\documentclass[sigconf]{acmart}
\usepackage{graphicx}
\usepackage{amsmath}
\usepackage{booktabs}
\usepackage{algorithm}
\usepackage{algorithmic}
\usepackage{amsfonts}
\usepackage{multirow}[c]
\usepackage{makecell}
\usepackage{subfigure}
\usepackage{color}
\usepackage{bm}
\usepackage{epstopdf}
\usepackage{url}
\usepackage[cal=cm]{mathalfa}
\usepackage{balance}
\usepackage{threeparttable}
\usepackage{wrapfig}
\usepackage{tabularx}
\usepackage{array}
\usepackage{enumitem}
\usepackage{appendix}
\usepackage{tikz}

\setlist[itemize]{leftmargin=*}

\makeatletter
\newcommand\notsotiny{\@setfontsize\notsotiny\@vipt\@viipt}
\makeatother

\AtBeginDocument{%
  }

\copyrightyear{2026}
\acmYear{2026}
\setcopyright{cc}
\setcctype{by}
\acmConference[WWW '26]{Proceedings of the ACM Web Conference 2026}{April 13--17, 2026}{Dubai, United Arab Emirates}
\acmBooktitle{Proceedings of the ACM Web Conference 2026 (WWW '26), April 13--17, 2026, Dubai, United Arab Emirates}
\acmPrice{}
\acmDOI{10.1145/3774904.3792189}
\acmISBN{979-8-4007-2307-0/2026/04}

\begin{document}

\title{S$^2$CDR: Smoothing-Sharpening Process Model for\\ Cross-Domain Recommendation}

\author{Xiaodong Li}
\affiliation{
  \institution{Institute of Information Engineering, Chinese Academy of Sciences \\School of Cyber Security, UCAS$^\dagger$}
  \city{Beijing, China} \\
  \country{lixiaodong@iie.ac.cn}
 }

\author{Juwei Yue}
\affiliation{
  \institution{Institute of Information Engineering, Chinese Academy of Sciences \\School of Cyber Security, UCAS}
  \city{Beijing, China} \\
  \country{yuejuwei@iie.ac.cn}
 }

\author{Xinghua Zhang}
\affiliation{
  \institution{Alibaba Group}
  \city{Beijing, China} \\
  \country{zxh.zhangxinghua@gmail.com}
 }

\author{Jiawei Sheng}
\affiliation{
  \institution{Institute of Information Engineering, Chinese Academy of Sciences}
  \city{Beijing, China} \\
  \country{shengjiawei@iie.ac.cn}
 }

\author{Wenyuan Zhang}
\affiliation{
  \institution{Institute of Information Engineering, Chinese Academy of Sciences \\School of Cyber Security, UCAS}
  \city{Beijing, China} \\
  \country{zhangwenyuan@iie.ac.cn}
 }

\author{Taoyu Su}
\affiliation{
  \institution{Institute of Information Engineering, Chinese Academy of Sciences}
  \city{Beijing, China} \\
  \country{sutaoyu@iie.ac.cn}
 }

\author{Zefeng Zhang}
\affiliation{
  \institution{Institute of Information Engineering, Chinese Academy of Sciences \\School of Cyber Security, UCAS}
  \city{Beijing, China} \\
  \country{zhangzefeng@iie.ac.cn}
 }

\author{Tingwen Liu}
\authornote{Corresponding author.\\ $^\dagger$University of Chinese Academy of Sciences.}
\affiliation{
  \institution{Institute of Information Engineering, Chinese Academy of Sciences \\School of Cyber Security, UCAS}
  \city{Beijing, China} \\
  \country{liutingwen@iie.ac.cn}
 }

\renewcommand{\shorttitle}{S$^2$CDR: Smoothing-Sharpening Process Model for Cross-Domain Recommendation}
\renewcommand{\shortauthors}{Xiaodong Li et al.}

\begin{abstract}
User cold-start problem is a long-standing challenge in recommendation systems. Fortunately, cross-domain recommendation (CDR) has emerged as a highly effective remedy for the user cold-start challenge, with recently developed diffusion models (DMs) demonstrating exceptional performance.
However, these DMs-based CDR methods focus on dealing with user-item interactions, overlooking correlations between items across the source and target domains. Meanwhile, the Gaussian noise added in the forward process of diffusion models would hurt user's personalized preference, leading to the difficulty in transferring user preference across domains.
To this end, we propose a novel paradigm of \textit{\textbf{S}moothing-\textbf{S}harpening Process Model} \textit{for \textbf{CDR}} to cold-start users, termed as \textbf{S$\bm{^2}$CDR} which features a corruption-recovery architecture and is solved with respect to ordinary differential equations (ODEs).
Specifically, the {\it smoothing process} gradually {\it corrupts} the original user-item/item-item interaction matrices derived from both domains into smoothed preference signals in a noise-free manner, and the {\it sharpening process} iteratively sharpens the preference signals to {\it recover} the unknown interactions for cold-start users.
Wherein, for the smoothing process, we introduce the heat equation on the item-item similarity graph to better capture the correlations between items across domains, and further build the tailor-designed low-pass filter to filter out the high-frequency noise information for capturing user's intrinsic preference, in accordance with the graph signal processing (GSP) theory.
Extensive experiments on three real-world CDR scenarios confirm that our S$\bm{^2}$CDR significantly outperforms previous SOTA methods in a training-free manner.

\end{abstract}

\ccsdesc[500]{Information systems~Recommender systems}

\keywords{Cross-Domain Recommendation, Cold-Start Recommendation, \\ Collaborative Filtering}

\maketitle

\section{Introduction}
Recommendation systems (RS)~\cite{tgt,ngcf,ncf} have been extensively applied in various online platforms, such as Alibaba (e-commerce) and TikTok (online video), leading to multiple user interactions in different categories on e-commerce websites and different media forms in online video. Consequently, developing effective strategies for providing recommendations to newly arrived (cold-start) users in various domains has emerged as a critical challenge in RS.

To alleviate the cold-start problem~\cite{cmclrec,pdma}, \textit{\textbf{cross-domain recommendation (CDR)}}~\cite{cdrnp,dmcdr,udmcf,farm} has received considerable attention in recent years, which aims to explore and exploit meaningful knowledge in related domains for achieving promising recommendation performance for cold-start users in another domain. 
A common idea of CDR methods is to utilize the overlapping users to bridge the source and target domains, and thus make recommendation for cold-start (non-overlapping) users who only have interactions in one domain.
Following the above idea, traditional CDR methods~\cite{emcdr,sscdr,udmcf} mostly adhere to the mapping-based paradigm, which encodes user/item representations in both domains separately, and then learns a mapping function to transfer user preference across domains. In addition, some meta-based methods~\cite{ptupcdr,cdrnp,remit} further adopt the meta-learning paradigm, treating each user's CDR as an individual task and learning mapping functions for each user to achieve personalized preference transfer.

Diffusion models (DMs)~\cite{ddpm} have achieved remarkable success in the field of RS~\cite{drm,dreamrec,difashion,diffkg,mcdrec} due to their representation reconstruction ability. More recently, DMCDR~\cite{dmcdr} pioneers the application of diffusion models to CDR, which gradually corrupts the user representation by adding Gaussian noise in the forward process, and then iteratively generates personalized user representation in the target domain under the guidance of user preference derived from user-item interactions in the source domain.
However, the DMs-based paradigm has the following two limitations for CDR task: (1) \textbf{Lack of Item Correlation}: the diffusion model emphasizes the user-item interactions, neglecting to capture the correlation between items across the source and target domains, which also serves as a crucial bridge for preference transfer. (2) \textbf{Detriment of External Noise}: the Gaussian noise added in the forward process would hurt user's intrinsic personalized preference, leading to transfer the sub-optimal user preference across domains.

To address the above issues, this paper proposes \textbf{S$\bm{^2}$CDR}, a novel \textit{\textbf{S}moothing-\textbf{S}harpening Process Model for \textbf{CDR}} to cold-start users, which possesses a corruption-recovery architecture. 
Specifically, the {\it smoothing process} gradually smooths the user/item information including both user-item and item-item interaction signals from both domains in a noise-free manner with tailor-designed graph filters, supported by graph signal processing (GSP)~\cite{gf-cf,pgsp,fire,higsp}. Wherein, we calculate the heat equation on the item-item similarity graph to capture the correlations between items across domains, and further build the effective low-pass filter to filter out the high-frequency information which is typically noise~\cite{pgsp,gf-cf}.
Afterwards, the {\it sharpening process} iteratively refines and sharpens preference signals to describe user's personalized preference, thus recovering the underlying interactions for cold-start users. 

To run our smoothing-sharpening process model, we further introduce ordinary differential equations (ODEs)~\cite{ltocf,node} for more effective and efficient cross-domain preference transfer via solving our smoothing and sharpening functions in a continuous manner, naturally characterizing the user's intrinsic preference for transfer and contributing to our training-free method.
Extensive experiments demonstrate that our S$^2$CDR outperforms previous SOTA methods in three real-world CDR scenarios. Meanwhile, as a non-parametric method, our S$^2$CDR shows high computational efficiency.
Our main contributions can be summarized as follows:
\begin{itemize}
    \item We inspire a corruption-recovery paradigm for CDR, which fulfills the bridging effect of item correlations in cross-domain preference transfer and retains user's intrinsic preference features.
    \item We propose a novel Smoothing-Sharpening Process Model S$^2$CDR for CDR to cold-start users, where the smoothing process smooths the user-item/item-item interaction signals for capturing the user preference based on tailor-designed graph filters, and the sharpening process sharpens the preference signals to recover unknown interactions for cold-start users.
    \item We conduct extensive experiments in three real-world CDR scenarios to demonstrate the effectiveness and efficiency of our S$^2$CDR over previous SOTA methods. 
    Despite the fact that S$^2$CDR does not require any training, we still achieve an average relative improvement of 7.07\%, 7.42\% with respect to HR and NDCG metrics over six cross-domain pairs.
\end{itemize}

\section{Preliminary}
In this section, we review the background knowledge of graph signal processing (GSP), ordinary differential equations (ODEs), and defines our CDR problem settings. We also review diffusion models (DMs) and give a formal comparison between DMCDR and our S$^2$CDR in \textbf{Appendix~\ref{diffusion_models}}.

\subsection{Graph Signal Processing}
In this subsection, we introduce basic concepts of graph signal processing (GSP)~\cite{gspml,gspoca}, including graph signal, graph filter, and graph convolution.

\subsubsection{\textbf{Graph signal}}
Given a weighted undirected graph $\mathcal{G} = (\mathcal{V},\mathcal{E})$, where $\mathcal{V} = \{v_1, v_2, \dots, v_n\}$ represents the vertex set with $n$ nodes (\textit{e.g.}, users or items), and $\mathcal{E}$ is the set of edges (\textit{e.g.}, user-item interactions). The graph $\mathcal{G}$ induces an adjacency matrix $\bm{A} \in \mathbb{R}^{n\times n}$, and if there is an edge between node $v_i$ and node $v_j$, then $\bm{A}_{i,j} = 1$, otherwise $\bm{A}_{i,j} = 0$. The graph signal is essentially a mapping $f : \mathcal{V} \to \mathbb{R}$, and it can be denoted as $\bm{x} = [x_1, x_2, \dots, x_n]^\top$, where $x_i$ represents the signal strength on node $v_i$.

Graph Laplacian matrix is the core concept in spectral graph theory~\cite{sgt} and it can be defined as $\bm{L}=\bm{D}-\bm{A}$, where $\bm{D}=\textit{diag}(\bm{A}\cdot\bm{1})$ is the degree matrix with $\bm{D}_{i,i}=\sum_{j}\bm{A}_{i,j}$. In addition, the normalized graph Laplacian matrix can be defined as $\tilde{\bm{L}}=\bm{I}-\tilde{\bm{A}}$, where $\bm{I}$ denotes the identity matrix and $\tilde{\bm{A}}=\bm{D}^{-1/2}\bm{A}\bm{D}^{-1/2}$.

\subsubsection{\textbf{Graph Filter and Graph Convolution}}
As the graph Laplacian matrix is real symmetric, it can be decomposed into $\bm{L}=\bm{U}\bm{\Lambda}\bm{U}^\top$, where $\bm{\Lambda}=\textit{diag}(\lambda_1,\lambda_2,\dots,\lambda_n)$ is the eigenvalue matrix with $\lambda_1\le\lambda_2\le\dots\le\lambda_n$, and $\bm{U}=(\bm{u}_1,\bm{u}_2,\dots,\bm{u}_n)$ is the eigenvector matrix with $\bm{u_i}\in\mathbb{R}^n$ being the eigenvector for  eigenvalue $\lambda_i$.

We can treat the eigenvector matrix $\bm{U}$ as the bases in Graph Fourier Transform (GFT). Specifically, given the graph signal $\bm{x}\in\mathbb{R}_n$, the GFT is defined as $\hat{\bm{x}}=\bm{U}^\top\bm{x}$, and the inverse transform is given by $\bm{x}=\bm{U}\hat{\bm{x}}$. Thus, the graph filter $\mathcal{H}(\bm{L})$ is defined as:
\begin{equation}
\begin{split}
\mathcal{H}(\bm{L})=\bm{U}\textit{diag}(h(\lambda_1),h(\lambda_2),\dots,h(\lambda_n))\bm{U}^\top,
\end{split}
\label{graph_filter}
\end{equation}
where $h(\cdot)$ is the filter defined on the eigenvalues. Given the input signal $\bm{x}$ and the filter $\mathcal{H}(\bm{L})$, the graph convolution is defined as:
\begin{equation}
\begin{split}
\bm{y}=\mathcal{H}(\bm{L})\bm{x}=\bm{U}\textit{diag}(h(\lambda_1),h(\lambda_2),\dots,h(\lambda_n))\bm{U}^{\top}\bm{x},
\end{split}
\label{}
\end{equation}
where the graph signal $\bm{x}$ is usually the observed user-item ratings~\cite{gf-cf,pgsp}, or the initial embeddings of users/items~\cite{lightgcn,ngcf} in CF. In summary, the original graph signal $\bm{x}$ is first transformed from spatial domain to spectral domain by Graph Fourier Transform basis $\bm{U}^\top$, then the spectral domain signal is processed by filter $h(\cdot)$, and finally the signal is transformed back to spatial domain through inverse Graph Fourier Transform basis $\bm{U}$.

\subsection{Ordinary Differential Equations}\label{ode_solvers}

The graph filters $\mathcal{H}(\bm{L})$ in Eq.~(\ref{graph_filter}) apply a discrete-time manner to handle the matrix $\bm{A}$. However, in order to capture the continuous dynamics of the graph structure, we set our sights to ordinary differential equations (ODEs)~\cite{ltocf,node} with a continuous-time manner.

Given the initial value $\bm{x}_0$ at time $t=0$, ODEs calculate $\bm{x}_T$ from $\bm{x}_0$ by solving the following Riemann integral problem~\cite{node}:
\begin{equation}
\begin{split}
\bm{x}_T=\bm{x}_0+\int_0^Tf(\bm{x}_t)dt,
\end{split}
\label{}
\end{equation}
where $f:\mathbb{R}^{dim(\bm{x})}\to\mathbb{R}^{dim(\bm{x})}$ is an ODE function\footnote{The ODE function $f$ can be learned through neural networks in the case of neural ordinary differential equations (NODEs)~\cite{node}. In contrast, we leverage the smoothing-sharpening function to approximate $f$ in this paper.} approximating the time-derivative of $\bm{x}$, \textit{i.e.}, $\frac{d\bm{x}_t}{dt}$.

Since it is usually infeasible to find an analytical solution for $\bm{x}_T$, we typically rely on existing ODE solvers, \textit{e.g.}, the explicit Euler method, the Dormand–Prince (DOPRI) method, the Runge-Kutta method and so on~\cite{erkf}. Specifically, ODE solvers discretize the time variable $t$ and transform the integral into a series of additions steps. The explicit Euler method can be written as follows:
\begin{equation}
\begin{split}
\bm{x}_{t+s}=\bm{x}_t+s\cdot f(\bm{x}_t),
\end{split}
\label{}
\end{equation}
where $s$ is a configured step size, which is usually smaller than 1. 

Other ODE solvers utilize more complicated method to update $\bm{x}_{t+s}$ from $\bm{x}_t$. For instance, the fourth-order Runge–Kutta (RK4) method can be expressed as follows:
\begin{equation}
\begin{split}
\bm{x}_{t+s}=\bm{x}_t+\frac{s}{6}(f_1+2f_2+2f_3+f_4),
\end{split}
\label{}
\end{equation}
where $f_1=f(\bm{x}_t)$, $f_2=f(\bm{x}_t+\frac{s}{2}f_1)$, $f_3=f(\bm{x}_t+\frac{s}{2}f_2)$, and $f_4=f(\bm{x}_t+sf_3)$.

It is also worth noting the implicit Adams-Moulto method, which can be derived as follows:
\begin{equation}
\begin{split}
\bm{x}_{t+s}=\bm{x}_t+\frac{s}{24}(9f_1+19f_2-5f_3+f_4),
\end{split}
\label{}
\end{equation}
where $f_1=f(\bm{x}_{t+s})$, $f_2=f(\bm{x}_t)$, $f_3=f(\bm{x}_{t-s})$, and $f_4=f(\bm{x}_{t-2s})$. The implicit method differs from the explicit methods above, as it makes use of past multi-step history (\textit{i.e.}, $f_3$ and $f_4$), which enables to learn a more robust derivative term.

To solve the above integral problem, we need to iterate the fixed-step ODE solvers $T/s$ times, and each iteration step updates $\bm{x}_t$ to $\bm{x}_{t+s}$. We consider all these ODE solvers in our experiments.

\begin{figure*}[t!]
\begin{center}
\includegraphics[width=15.5cm]{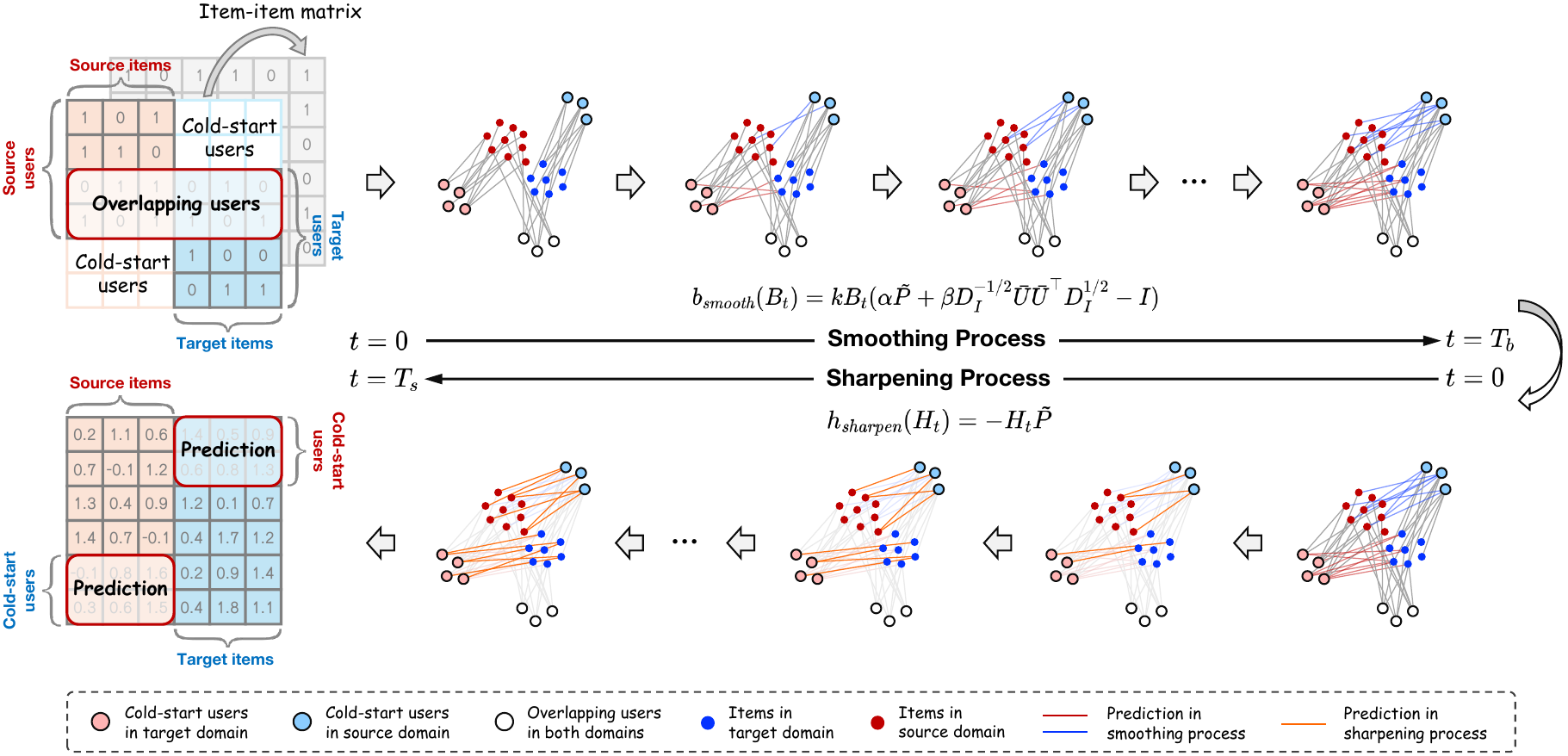}
\caption{The overall framework of S$^2$CDR. The smoothing process aims to gradually corrupt the original user-item/item-item interaction signals into smoothed preference signals in a noise-free manner, while the sharpening process iteratively refines and sharpens the preference signals to recover the unknown interactions for cold-start users.}
\label{main_model}
\end{center}
\end{figure*}

\subsection{Problem Setting}\label{problem_setting}
We assume there are two domains, \textit{i.e.}, a source domain $\mathcal{D}^s$ and a target domain $\mathcal{D}^t$. Formally, the user-item interaction data can be denoted as $\mathcal{D}^s=\{\mathcal{U}^s,\mathcal{V}^s,\mathcal{Y}^s\}$, and $\mathcal{D}^t=\{\mathcal{U}^t,\mathcal{V}^t,\mathcal{Y}^t\}$, where $\mathcal{U}$, $\mathcal{V}$ and $\mathcal{Y}$ represent user set, item set and rating scores respectively. We define the user-item interaction matrix as $\bm{R}^s\in\mathbb{R}^{|\mathcal{U}^s|\times|\mathcal{V}^s|}$ and $\bm{R}^t\in\mathbb{R}^{|\mathcal{U}^t|\times|\mathcal{V}^t|}$ in the source and target domains, where $\bm{R}_{i,j}=1$ indicates that user $u_i$ has interacted with item $v_j$, otherwise $\bm{R}_{i,j}=0$. To simplify the problem setting, we assume both domains have no other auxiliary information. Furthermore, we define the set of overlapping users that exist in both domains as $\mathcal{U}^o=\mathcal{U}^s\cap\mathcal{U}^t$. Consequently, the cold-start users that exist only in the source domain can be denoted as $\mathcal{U}^{s\setminus o} = \mathcal{U}^{s}\setminus \mathcal{U}^o$, as well as cold-start users $\mathcal{U}^{t\setminus o} = \mathcal{U}^{t}\setminus \mathcal{U}^o$ that exist only in the target domain.

Given the user-item interaction matrix (\textit{e.g.}, $\bm{R}^s$ and $\bm{R}^t$) of overlapping users and cold-start (non-overlapping) users in both the source and target domains, our CDR goal is to predict the items that the cold-start users may interact with in the source or target domain using the cross-domain interaction patterns of overlapping users.
To achieve the above goal, following previous methods~\cite{emcdr,ptupcdr,udmcf,dmcdr}, we employ a unified user-item interaction matrix $\bm{R}^{st}\in\mathbb{R}^{(|\mathcal{U}^o|+|\mathcal{U}^{s\setminus o}|+|\mathcal{U}^{t\setminus o}|)\times(|\mathcal{V}^s|+|\mathcal{V}^t|)}$, which leverages the overlapping users to bridge the source and target domains as follows:
\begin{equation}
\begin{split}
\bm{R}^{st}=
\begin{bmatrix}
\mathbb{R}^{|\mathcal{U}^{s\setminus o}|\times|\mathcal{V}^s|} & \mathbb{R}^{|\mathcal{U}^{s\setminus o}|\times|\mathcal{V}^t|} \\
\mathbb{R}^{|\mathcal{U}^o|\times|\mathcal{V}^s|} & \mathbb{R}^{|\mathcal{U}^o|\times|\mathcal{V}^t|} \\
\mathbb{R}^{|\mathcal{U}^{t\setminus o}|\times|\mathcal{V}^s|} &
\mathbb{R}^{|\mathcal{U}^{t\setminus o}|\times|\mathcal{V}^t|}
\end{bmatrix},
\end{split}
\label{R_st}
\end{equation}
where $\mathbb{R}^{|\mathcal{U}^{s\setminus o}|\times|\mathcal{V}^s|}$, $\mathbb{R}^{|\mathcal{U}^o|\times|\mathcal{V}^s|}$, $\mathbb{R}^{|\mathcal{U}^o|\times|\mathcal{V}^t|}$ and $\mathbb{R}^{|\mathcal{U}^{t\setminus o}|\times|\mathcal{V}^t|}$ are the observed user-item interaction matrix.
Meanwhile, $\mathbb{R}^{|\mathcal{U}^{s\setminus o}|\times|\mathcal{V}^t|}$ and $\mathbb{R}^{|\mathcal{U}^{t\setminus o}|\times|\mathcal{V}^s|}$ are all-zero matrix with cold-start users, which will be inferred by our model.

To ensure equal impact from nodes (users/items) of varying degrees, we define the normalized interaction matrix as $\tilde{\bm{R}}^{st}=\bm{D}^{-1/2}_U\bm{R}^{st}\bm{D}^{-1/2}_I$, where $\bm{D}^{-1/2}_U=\textit{diag}(\bm{R}^{st}\cdot\bm{1})$ and $\bm{D}^{-1/2}_I=\textit{diag}(\bm{1}^{\top}\bm{R}^{st})$ are the degree matrices of $\bm{R}^{st}$. We further define the normalized adjacency matrix of the item-item similarity graph as $\tilde{\bm{P}}={\tilde{\bm{R}}^{st\top}}\tilde{\bm{R}}^{st}$.
The notations used in this paper are summarized in \textbf{Appendix~\ref{notations}}.

\section{Methodology}
This section introduces our proposed S$^2$CDR implementing the smoothing-sharpening principle, as described in Figure~\ref{main_model}. Our S$^2$CDR mainly consists of two parts, \textit{i.e.}, the \textit{smoothing process} and the \textit{sharpening process}. The smoothing process aims to corrupt the original user-item interaction matrix into smoothed preference signals in a noise-free manner. As for the recovery stage, the sharpening process iteratively refines and sharpens preference signals to generate new interactions for cold-start users in both domains.

\subsection{Overview}
 
Given the user-item interaction matrix $\bm{R}^{st}$ (described in Eq.~(\ref{R_st})), we first apply a continuous smoothing process to $\bm{R}^{st}$ to derive its corrupted matrix $\bm{B}_{T_b}$, which contains smoothed preference signals. Subsequently, we introduce a continuous sharpening process to $\bm{B}_{T_b}$ to derive its refined and sharpened matrix $\bm{H}_{T_h}$, which can reveal hidden interactions for cold-start users.

\textit{Analysis of the smoothing process}. The smoothing process is the core of collaborative filtering, and has been widely explored as graph convolutional filters in many GSP-based methods~\cite{fire,higsp,pgsp,gf-cf,cgsp}. In these traditional methods, the candidate items are recommended to users after this process.

\textit{Analysis of the sharpening process}. The sharpening process is used to refine and sharpen the preference signals obtained by the smoothing process. Taking one step further, the sharpening process is designed to describe user's personalized preference. Similar to the reverse process of DMs~\cite{ddpm}, unknown user-item interactions are predicted for cold-start users after this process.

We also further emphasize that our S$^2$CDR is a non-parametric approach with high computational efficiency. During the smoothing-sharpening process, our model can be obtained in a closed-form solution instead of expensive back-propagation training,
as our S$^2$CDR directly handles the user-item/item-item interaction matrix and there's no need for neural networks or user/item embeddings.

\subsection{Smoothing Process}
In the context of CDR, our goal is to predict the potential items that the cold-start users may interact with in both domains, which is only relevant to user-item/item-item interaction matrix $\bm{R}^{st}$/$\tilde{\bm{P}}$. Therefore, we argue that it is neither efficient nor necessary to add Gaussian noise to corrupt the interaction matrix, as is done in the forward process of DMs. In addition, adding too much noise to interaction matrix would also destroy user's personalized preference modeling. To this end, we introduce a noise-free manner dedicated to graph signals, which repeats the smoothing operation until the interaction matrix reaches a steady state.

\subsubsection{ODE-based Smoothing Process}
The general idea of smoothing process is to solve the Riemann integral problem to derive the smoothed preference signals of cold-start users, which can be formulated as follows:
\begin{equation}
\begin{split}
\bm{B}_{T_b}=\bm{B}_0+\int_0^{T_b}b(\bm{B}_t)dt,
\end{split}
\label{smoothing_process}
\end{equation}
where the input $\bm{B}_0$ is user-item interaction matrix $\bm{R}^{st}$, $T_b$ is the terminal time, and $b:\mathbb{R}^{dim(B)}\to\mathbb{R}^{dim(B)}$ is the smoothing function which describes the time-derivative of $\bm{R}^{st}$, \textit{i.e.}, $\frac{d\bm{B}_t}{dt}$.

The key to the smoothing process lies in the definition of the smoothing function $b$, which we will describe shortly.

\subsubsection{Smoothing with Heat Equation}
The heat equation describes the law of thermal diffusive processes, \textit{i.e.}, Newton's Law of Cooling. This concept has been used in image synthesis~\cite{bdm,gmihd} for blurring images with blurring filters. Inspired by this, we leverage the heat equation to capture the correlation between items from $\tilde{\bm{P}}$ as follows:
\begin{equation}
\begin{split}
b_{\textit{heat}}(\bm{B}_t)=k\bm{B}_t(\tilde{\bm{P}}-\bm{I}),
\end{split}
\label{heat}
\end{equation}
where $k\in\mathbb{R}$ is a hyper-parameter in our model, called the heat capacity. The definition of the smoothing function $b_{\textit{heat}}$ is similar to the linear filters in GSP-based CF methods~\cite{pgsp,gf-cf,higsp}, which can capture the item-item similarity across the source and target domains and serve as a crucial bridge for preference transfer.

\subsubsection{Smoothing with Ideal Low-pass Filter}
The smoothing function $b_{\textit{heat}}$ aims to capture the low-order information of the graph. Inspired by the graph convolutions, we strengthen the expressiveness of the smoothing function with the ideal low-pass filter to obtain the high-order information in $\bm{R}^{st}$ as follows:
\begin{equation}
\begin{split}
b_{\textit{ideal}}(\bm{B}_t)=\bm{B}_t(\bm{D}^{-1/2}_I\bar{\bm{U}}\bar{\bm{U}}^{\top}\bm{D}^{1/2}_I-\bm{I}),
\end{split}
\label{ideal}
\end{equation}
where $\bar{\bm{U}}$ is the top-K singular vectors of $\bm{R}^{st}$. The smoothing function $b_{\textit{heat}}$ filters out the noise signals dominated by high-frequency information and preserves the user's intrinsic preference signals dominated by low-frequency information.

We combine the heat equation $b_{\textit{heat}}$ and the ideal low-pass filter $b_{\textit{ideal}}$ to define the final smoothing function as follows:
\begin{equation}
\begin{split}
b_{\textit{smooth}}(\bm{B}_t)&=\alpha\cdot b_{\textit{heat}}(\bm{B}_t)+\beta\cdot b_{\textit{ideal}}(\bm{B}_t)\\
&=k\bm{B}_t(\alpha\tilde{\bm{P}}+\beta\bm{D}^{-1/2}_I\bar{\bm{U}}\bar{\bm{U}}^{\top}\bm{D}^{1/2}_I-\bm{I}),
\end{split}
\label{smoothing_function}
\end{equation}
where $k=1$, $\alpha$ and $\beta$ are hyper-parameters that controls the strength of $b_{\textit{heat}}$ and $b_{\textit{ideal}}$. 

As a result, the smoothing process in Eq.~(\ref{smoothing_process}) can be written as follows with smoothing function $b_{\textit{smooth}}$ in Eq.~(\ref{smoothing_function}):
\begin{equation}
\begin{split}
\bm{B}_{T_b}=\bm{B}_0+\int_0^{T_b}b_{\textit{smooth}}(\bm{B}_t)dt,
\end{split}
\label{smoothing_process_final}
\end{equation}
where $\bm{B}_{T_b}$ is the user-item interaction matrix with smoothed preference signals, and Eq.~(\ref{smoothing_process_final}) can be inferred by the ODE solvers we introduced in Section~\ref{ode_solvers}.

\subsection{Sharpening Process}
As the counterpart to the smoothing process, we expect the sharpening process to iteratively refine and sharpen the smoothed preference signals to emphasize the differences between nodes, thereby describing user's personalized preference, and ultimately generating new interactions for cold-start users in both domains. The sharpening process can be written as follows:
\begin{equation}
\begin{split}
\bm{H}_{T_h}=\bm{H}_0+\int_0^{T_h}h(\bm{H}_t)dt,
\end{split}
\label{}
\end{equation}
where the input matrix $\bm{H}_0=\bm{B}_{T_b}$ is the output of the smoothing process in Eq.~(\ref{smoothing_process_final}), $T_h$ is the terminal time, and $h:\mathbb{R}^{dim(H)}\to\mathbb{R}^{dim(H)}$ is the sharpening function which approximates $\frac{d\bm{H}_t}{dt}$.

In order to emphasize the differences from the neighbors, and thus capture user's personalized preference, we define the sharpening function as follows:
\begin{equation}
\begin{split}
h_{\textit{sharpen}}(\bm{H}_t)=-\bm{H}_t\tilde{\bm{P}},
\end{split}
\label{sharpen_function}
\end{equation}
where the negative sign is added to refine and sharpen the smoothed preference signals. Consequently, the final sharpening process can be formulated as follows:
\begin{equation}
\begin{split}
\hat{\bm{R}}^{st}=\bm{H}_0+\int_0^{T_h}h_{\textit{sharpen}}(\bm{H}_t)dt,
\end{split}
\label{sharpening_process}
\end{equation}
where $\bm{H}_0=\bm{B}_{T_b}$, and $\hat{\bm{R}}^{st}$ is the sharpened user-item interaction matrix which contains user's personalized preference to infer unknown user-item interactions for cold-start users in $\mathbb{R}^{|\mathcal{U}^{s\setminus o}|\times|\mathcal{V}^t|}$ and $\mathbb{R}^{|\mathcal{U}^{t\setminus o}|\times|\mathcal{V}^s|}$. Similarly, Eq.~(\ref{sharpening_process}) can be solved by ODE solvers mentioned in Section~\ref{ode_solvers}.

\subsection{Training-Free Procedure}
It is worth noting that our S$^2$CDR is non-parametric, which means there is no training phase and it has very high computational efficiency. Since our model directly handles the user-item/item-item interaction matrix $\bm{R}^{st}$/$\tilde{\bm{P}}$, we do not need neural networks to learn user/item embeddings. In addition, we introduce the ODE solvers to infer Eq.~(\ref{smoothing_process_final}) and Eq.~(\ref{sharpening_process}) in a continuous-time manner, which further reduces the overall computational time. Meanwhile, our S$^2$CDR shows the best performance compared to other SOTA baselines.
The overall procedure of S$^2$CDR is shown in \textbf{Appendix~\ref{alg1}}.

\begin{table*}[t]
\footnotesize
\centering
\caption{Overall performance comparison between the baselines and our S$\bm{^2}$CDR in three CDR scenarios. The bold results highlight the best results, while the second-best results are underlined. \% Improve represents the relative improvements of our S$\bm{^2}$CDR over the best baseline.}
\label{mainexperiment}
\setlength\tabcolsep{2.0pt}
\begin{tabular*}{1 \textwidth}
{@{\extracolsep{\fill}}@{}lcccccccccccc@{}}
\toprule
&
\multicolumn{4}{c}{\textbf{Douban:~}\textit{\textbf{Movie}}~$\&$~\textit{\textbf{Book}}} & \multicolumn{4}{c}{\textbf{Amazon:~}\textit{\textbf{Movie}}~$\&$~\textit{\textbf{Music}}} & \multicolumn{4}{c}{\textbf{Amazon:~}\textit{\textbf{Book}}~$\&$~\textit{\textbf{Music}}}    \\
\cmidrule(r){2-5}\cmidrule(r){6-9}\cmidrule(r){10-13} \bf Methods &
\multicolumn{2}{c}{ \bf{Book~$\to$~Movie}} & \multicolumn{2}{c}{ \bf{Movie~$\to$~Book}} &  \multicolumn{2}{c}{ \bf{Music~$\to$~Movie}} & \multicolumn{2}{c}{ \bf{Movie~$\to$~Music}} &  \multicolumn{2}{c}{ \bf{Music~$\to$~Book}} & \multicolumn{2}{c}{ \bf{Book~$\to$~Music}} \\
\cmidrule(r){2-3}\cmidrule(r){4-5}\cmidrule(r){6-7}
\cmidrule(r){8-9}\cmidrule(r){10-11}\cmidrule(r){12-13}&
\bf{HR} & \bf{NDCG} & \bf{HR} & \bf{NDCG} & \bf{HR} & \bf{NDCG} & \bf{HR} & \bf{NDCG} & \bf{HR} & \bf{NDCG} & \bf{HR} & \bf{NDCG} \\
\midrule
NeuMF  &   0.2390   &   0.1441   &   0.2276   &   0.1105     & 
 0.1530   &   0.0929   &   0.1581   &   0.0883    &   0.1395   &   0.0707   &   0.1518   &   0.0781   \\

EMCDR     & 0.2477   &   0.1525 &  0.2342   &    0.1193  &    0.1569    &    0.0993  &  0.1662  &  0.0978  &  0.1481  &  0.0860  &  0.1621  &  0.0898  \\
DCDCSR &  0.2519   &   0.1560 &  0.2409   &    0.1221  &    0.1613    &    0.1028  &  0.1688  &  0.1026  &  0.1513  &  0.0889  &  0.1634  &  0.0935  \\
SSCDR   & 0.2558   &   0.1547 &  0.2462   &    0.1264  &    0.1639    &    0.1051  &  0.1726  &  0.1012  &  0.1544  &  0.0920  &   0.1673  &  0.0972  \\
TMCDR  &  0.2602 &  0.1596   &  0.2514 &  0.1285 &  0.1672 &  0.1089  &  0.1763  &  0.1041  &  0.1576  &  0.0916  &  0.1696  &  0.1009 \\
LACDR   &  0.2643 &  0.1615   &  0.2527 &  0.1298 &  0.1683 &  0.1104  &  0.1775  &  0.1060  &  0.1594  &  0.0935  &  0.1710  &  0.1022  \\
DOML   &   0.2667 &  0.1642   &  0.2553 &  0.1320 &  0.1714 &  0.1122  &   0.1795  &  0.1087  &  0.1608  &  0.0952  &  0.1715  &  0.1036 \\
BiTGCF  &   0.2703 &  0.1631   &  0.2578 &  0.1349 &  0.1693 &  0.1107  &   0.1792  &  0.1076  &  0.1637  &  0.0978  &  0.1743  &  0.1055 \\
PTUPCDR  &   0.2736 &  0.1634   &  0.2590 &  0.1356 &  0.1709 &  0.1145  &  0.1810  &  0.1094  &  0.1622  &  0.0975  &  0.1749  &  0.1067 \\
CDRIB  &  0.2761   &  0.1662 &  0.2628 &  0.1375 &  0.1718  &  0.1140  &  0.1832  &  0.1123  &  0.1646  &  0.0989  &  0.1758  &  0.1069 \\
UDMCF  &  0.2914   &  0.1792 &  0.2778 &  0.1543 &  0.1887  &  0.1304  &  0.1952  &  0.1236  &  \underline{0.1756}  &  \underline{0.1112}  &  0.1863  &  0.1174 \\
GF-CF  &  0.4171   & 0.2168  & 0.2580  & 0.1566  &  0.2677  &  0.1813  & 0.2097   &  0.1486  &  0.0961  &  0.0481  &  0.1700  & 0.0859  \\
PGSP  &  0.4301   & 0.2470  & \underline{0.2956}  & \underline{0.1734}  &  0.2710  &  0.1851  &  0.2163  &  0.1516  &  0.1003  &  0.0500  &  0.1736  &  0.0881 \\
DMCDR  &  \underline{0.4925}   & \underline{0.3046}  & 0.2825  & 0.1624  &  \underline{0.3515}  &  \underline{0.2169}  &  \underline{0.2968}  &  \underline{0.1920}  &  0.1662  &  0.1070  &  \underline{0.2643}  & \underline{0.1510}  \\
\midrule
S$^2$CDR  &  \textbf{0.5430}  &  \textbf{0.3302}  &  \textbf{0.3172}  &  \textbf{0.1813}  &  \textbf{0.3703}  &  \textbf{0.2420}  &  \textbf{0.3139}  &  \textbf{0.2049}  &  \textbf{0.1905}  &  \textbf{0.1159}  &  \textbf{0.2782}  &  \textbf{0.1646} \\
{\%} Improve & 10.25{\%}  &  8.40{\%}  &  7.31{\%}  &  4.56{\%}  &  5.35{\%}  &  11.57{\%}  & 5.76{\%}  &  6.72{\%}  &  8.49{\%}  &  4.23{\%}  &  5.26{\%}  &  9.01{\%}\\
\bottomrule
\end{tabular*}
\end{table*}

\section{Experiments}

In this section, we conduct extensive experiments on several real-world datasets to answer the following questions:
\begin{itemize}
    \item \textbf{RQ1:} How does our S$^2$CDR perform compared to state-of-the-art baselines, including single-domain and cross-domain methods?
    \item \textbf{RQ2:} How do the smoothing and sharpening functions benefit the performance of S$^2$CDR?
    \item \textbf{RQ3:} Does the heat equation proposed in the smoothing process really capture the correlations between items?
    \item \textbf{RQ4:} Does the ideal low-pass filter proposed in the smoothing process really filter out the noise information?
    \item \textbf{RQ5:} How does the computational efficiency of our S$^2$CDR?
\end{itemize}

\subsection{Experimental Settings}

\subsubsection{\textbf{Datasets}}
Following previous works~\cite{tmcdr}, we conduct extensive experiments on two popular real-world datasets, i.e., \textbf{Douban} and \textbf{Amazon}. We select two relevant domains from Douban, including Movie and Book as scenario 1 : \textit{Movie}~$\&$~\textit{Book}. Similarly, we select three relevant domains from Amazon, including Movie and TV (termed as \textit{Movie}), CDs and Vinyl (termed as \textit{Music}), Books (termed as \textit{Book}) to form scenario 2 : \textit{Movie}~$\&$~\textit{Music} and scenario 3 : \textit{Book}~$\&$~\textit{Music}. Both datasets contain rating scores from 1 to 5, reflecting user's preference for specific items. As we consider implicit feedback in this paper, we binarize the ratings to 0 and 1. Specifically, we take the rating scores higher or equal to 4 as 1 and the others as 0. For each CDR scenario, we filter out the users with fewer than 5 interactions as previous works~\cite{tmcdr,cdrib}. The details of three CDR scenarios are summarized in \textbf{Appendix~\ref{dataset}}.

\subsubsection{\textbf{Baselines}}
We compare our \textbf{S$\bm{^2}$CDR} with the following SOTA baselines: (1) \textbf{NeuMF}~\cite{ncf}, (2) \textbf{EMCDR}~\cite{emcdr}, (3) \textbf{DCDCSR}~\cite{dcdcsr}, (4) \textbf{SSCDR}~\cite{sscdr}, (5) \textbf{TMCDR}~\cite{tmcdr}, (6) \textbf{LACDR}~\cite{lacdr}, (7) \textbf{DOML}~\cite{doml}, (8) \textbf{BiTGCF}~\cite{bitgcf}, (9) \textbf{PTUPCDR}~\cite{ptupcdr}, (10) \textbf{CDRIB}~\cite{cdrib}, (11) \textbf{UDMCF}~\cite{udmcf}, (12) \textbf{GF-CF}~\cite{gf-cf}, (13) \textbf{PGSP}~\cite{pgsp}, and (14) \textbf{DMCDR}~\cite{dmcdr}. The details of these baselines are shown in \textbf{Appendix~\ref{baselines}}.

\subsubsection{\textbf{Evaluation Protocol}}
Following previous studies~\cite{cdrib,udmcf,sscdr,tmcdr}, we evaluate our model and baselines with widely used \textit{leave-one-out} evaluation method: only 1 positive item and 999 randomly selected negative items. We adopt two evaluation metrics of predictive performance: HR@$k$ (Hit Rate) and NDCG@$k$ (Normalized Discounted Cumulative Gain) with $k = 10$.

\subsubsection{\textbf{Implementation Details}}
We implement the above methods using PyTorch 1.9.0 with python 3.6.
To keep the comparison fair with previous methods, we tune the hyper-parameters of each method according to their original literature.
For our S$^2$CDR, the ODE solvers to solve the integral problems of the smoothing and sharpening processes is chosen from Euler, RK4 and DOPRI. The size of $\alpha$ and $\beta$ is searched in $\left\{{0.1, 0.2, \cdots, 1.0}\right\}$. For the smoothing process, the terminal time $T_b$ is set to 1 to 3, and the number of steps $\frac{T_b}{s}$ is in the range of $\left\{{1, 2, 3, 4, 5}\right\}$. For the sharpening process, the terminal time $T_h$ is set to 1 to 3, and the number of steps $\frac{T_h}{s}$ is chosen from $\left\{{1, 2, 3, 4, 5}\right\}$. Besides, the heat capacity $k$ is set to 1.
The best hyper-parameter configurations in each CDR scenario are summarized in \textbf{Appendix~\ref{best_hyperparameters}}.

Following previous works~\cite{ptupcdr,cdrnp,dmcdr}, we randomly select about 20\% overlapping users as cold-start users (\textit{e.g.}, 10\% from Movie to make recommendation in Book, and the residual 10\% from Book to make recommendation in Movie), while the samples of other overlapping users are used to bridge the source and target domains.

\subsection{Overall Performance (RQ1)}
We compare our S$^2$CDR with other state-of-the-art baselines in three CDR scenarios. From the results shown in Table~\ref{mainexperiment}, we can summarize the following observations.

Compared to the single-domain method (\textit{e.g.}, NeuMF), cross-domain methods show a superior performance, highlighting the importance of transferring valuable user preference across domains.
Moreover, we also observe that our S$^2$CDR consistently outperforms the CF-based methods (\textit{e.g.}, PGSP). Such an observation suggests that S$^2$CDR is capable of describing user's personalized preference by employing the sharpening process to refine and sharpen the smoothed preference signals, as these CF-based methods can be considered to be based on the smoothing process only. Besides, while the mapping-based methods (\textit{e.g.}, EMCDR) and the meta-based methods (\textit{e.g.}, PTUPCDR) utilize mapping functions to transfer user preference across domains, we propose a novel smoothing-sharpening paradigm to directly handle the user-item/item-item interaction matrix, which outperforms these methods by a large margin, indicating the superiority of our methods in solving the cold-start issue. Furthermore, we find that S$^2$CDR can effectively improve the recommendation performance compared to the DMs-based method (\textit{e.g.}, DMCDR). Such improvements are attributed to the graph filters we designed in the smoothing process, which enable our model to capture the correlations between items and filter out the high-frequency noise information. Overall, S$^2$CDR consistently outperforms all the baselines in all CDR scenarios, which can be attributed to the following reasons: (1) the smoothing process to smooth the preference signals in a noise-free manner, (2) the sharpening process to refine and sharpen the preference signals, and (3) the ODE solvers to solve the smoothing and sharpening functions in a continuous-time manner. We will analyze the effectiveness of each component in later sections.

\begin{table}[t]
\footnotesize
\centering
\caption{Ablation Study in three CDR scenarios.}
\label{ablation}
\setlength\tabcolsep{5pt}
\begin{tabular*}{0.45 \textwidth}{@{\extracolsep{\fill}}@{}cc|cccc|c@{}}
\toprule
\multirow{2}{*}{\bf Scenarios}  &  \multirow{2}{*}{\bf Metric}  &  \textit{w}/\textit{o}  &  \textit{w}/\textit{o}  &  \textit{w}/\textit{o}  & \textit{w}/\textit{o}  &  \multirow{2}{*}{S$^2$CDR}  \\
&  & $b_{\textit{heat}}$ & $b_{\textit{ideal}}$ & $b_{\textit{smooth}}$ & $h_{\textit{sharpen}}$ & \\
\midrule
\multirow{2}{*}{Book~$\to$~Movie}  &  HR  &  0.3387  &  0.4995  & 0.1935  &  0.4892  &  \textbf{0.5430}  \\
& NDCG  &  0.1959  &  0.2979  &  0.0920  &  0.2741  & \textbf{0.3302} \\
\multirow{2}{*}{Movie~$\to$~Book}  &  HR  &  0.2043  &  0.3010  &  0.0645  &  0.3010  & \textbf{0.3172} \\
& NDCG  &  0.1446  &  0.1759  &  0.0205  &  0.1755  & \textbf{0.1813} \\
\midrule
\multirow{2}{*}{Music~$\to$~Movie}  &  HR  & 0.3424  &  0.2983  &  0.1444  &  0.3441  & \textbf{0.3703} \\
& NDCG  &  0.2069  &  0.1953  &  0.0767  &  0.2245  & \textbf{0.2420} \\
\multirow{2}{*}{Movie~$\to$~Music}  &  HR  &  0.2855  &  0.2230  &  0.1115  &  0.2973  & \textbf{0.3139} \\
& NDCG  &  0.1748  &  0.1568  &  0.0583  &  0.1913  & \textbf{0.2049} \\
\midrule
\multirow{2}{*}{Music~$\to$~Book}  &  HR  &  0.1668  &  0.1099  &  0.0823  &  0.1729  & \textbf{0.1905} \\
& NDCG  &  0.0943  &  0.0529  &  0.0412  &  0.1056  & \textbf{0.1159} \\
\multirow{2}{*}{Book~$\to$~Music}  &  HR  &  0.2370  &  0.1616  &  0.1370  &  0.2684  & \textbf{0.2782} \\
& NDCG  &  0.1337  &  0.0940  &  0.0798  &  0.1591  & \textbf{0.1646} \\
\bottomrule
\end{tabular*}
\vspace{-0.3cm}
\end{table}

\subsection{Ablation Study (\textbf{RQ2})}
To evaluate how does each component of S$^2$CDR contributes to the final performance, we conduct an ablation study in three CDR scenarios to compare our S$^2$CDR with four distinct variants as:
\begin{itemize}
    \item \textit{w}/\textit{o} $b_{\textit{heat}}$ : This variant removes the heat equation $b_{\textit{heat}}$.
    \item \textit{w}/\textit{o} $b_{\textit{ideal}}$ : This variant excludes the ideal low-pass filter $b_{\textit{ideal}}$.
    \item \textit{w}/\textit{o} $b_{\textit{smooth}}$ : We remove the smoothing function $b_{\textit{smooth}}$ in our model, which is equivalent to removing both $b_{\textit{heat}}$ and $b_{\textit{ideal}}$.
    \item \textit{w}/\textit{o} $b_{\textit{sharpen}}$ : This variant indicates the removal of the sharpening function $b_{\textit{sharpen}}$.
\end{itemize}
The comparison results are shown in Table~\ref{ablation}. We observe that S$^2$CDR performs significantly outperforms \textit{w}/\textit{o} $b_{\textit{heat}}$ variant, verifying the effectiveness of the heat equation $b_{\textit{heat}}$, which aims to capture correlations between items across domains. Meanwhile, we also find that the removal of $b_{\textit{ideal}}$ leads to performance degradation. This observation demonstrate the importance of using the ideal low-pass filter to filter out the high-frequency noise information. Another observation is that the performance decreases the most when $b_{\textit{smooth}}$ is removed, indicating the effectiveness of the customized graph filters we designed in the smoothing process, and we should combine $b_{\textit{heat}}$ and $b_{\textit{ideal}}$ for better recommendation performance. Lastly, compared with $b_{\textit{sharpen}}$ variant, S$^2$CDR further improves the performance. This finding validates that the sharpening process is indispensable in our model to describe user's personalized preference.

\begin{table}[t!]
\footnotesize
\centering
\caption{Item-item similarity analysis in three CDR scenarios.}
\label{item_item}
\setlength\tabcolsep{5pt}
\begin{tabular*}{0.38 \textwidth}{@{\extracolsep{\fill}}@{}cc|cc|c@{}}
\toprule
\bf Scenarios  &  \bf Metric  &  \textit{w}/ $\tilde{\bm{P}}_{\textit{source}}$  &  \textit{w}/ $\tilde{\bm{P}}_{\textit{target}}$  &  S$^2$CDR  \\
\midrule
\multirow{2}{*}{Book~$\to$~Movie}  &  HR  &  0.2526   &  0.3064  &  \textbf{0.5430}  \\
& NDCG  &  0.1583  &  0.1806  & \textbf{0.3302} \\
\multirow{2}{*}{Movie~$\to$~Book}  &  HR  &  0.2580  &  0.2634  & \textbf{0.3172} \\
& NDCG  &  0.1601  &  0.1595  & \textbf{0.1813} \\
\midrule
\multirow{2}{*}{Music~$\to$~Movie}  &  HR  & 0.1645  &  0.3535  & \textbf{0.3703} \\
& NDCG  &  0.0911  &  0.2174  & \textbf{0.2420} \\
\multirow{2}{*}{Movie~$\to$~Music}  &  HR  &  0.3100  &  0.1260  & \textbf{0.3139} \\
& NDCG  &  0.1924  &  0.0685  & \textbf{0.2049} \\
\midrule
\multirow{2}{*}{Music~$\to$~Book}  &  HR  &  0.0510  &  0.1829  & \textbf{0.1905} \\
& NDCG  &  0.0330  &  0.1056  & \textbf{0.1159} \\
\multirow{2}{*}{Book~$\to$~Music}  &  HR  &  0.2648  &  0.1382  & \textbf{0.2782} \\
& NDCG  &  0.1606  &  0.0763   & \textbf{0.1646} \\
\bottomrule
\end{tabular*}
\end{table}

\begin{figure}[t]
\setlength{\abovecaptionskip}{0.cm}
	\begin{center}
        \subfigure
        {\begin{minipage}[b]{.32\linewidth}
        \centering
        \includegraphics[scale=0.265]{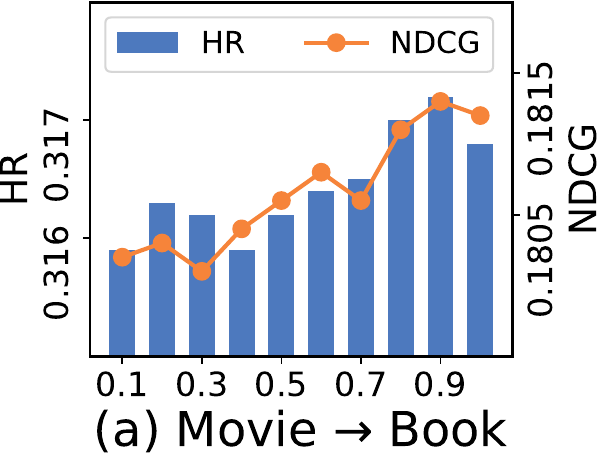}
        \end{minipage}}
        \subfigure
        {\begin{minipage}[b]{.32\linewidth}
        \centering
        \includegraphics[scale=0.265]{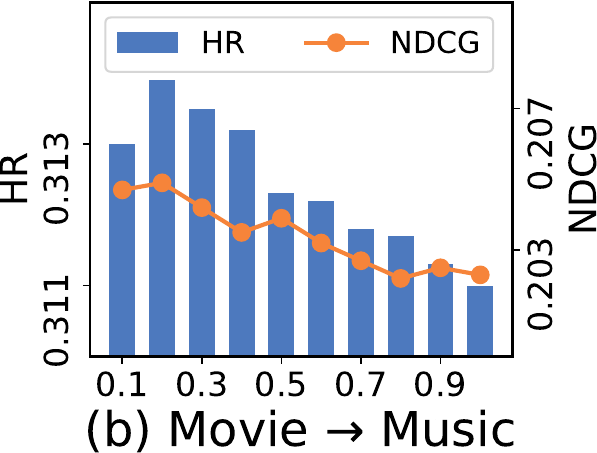}
        \end{minipage}}
        \subfigure
        {\begin{minipage}[b]{.32\linewidth}
        \centering
        \includegraphics[scale=0.265]{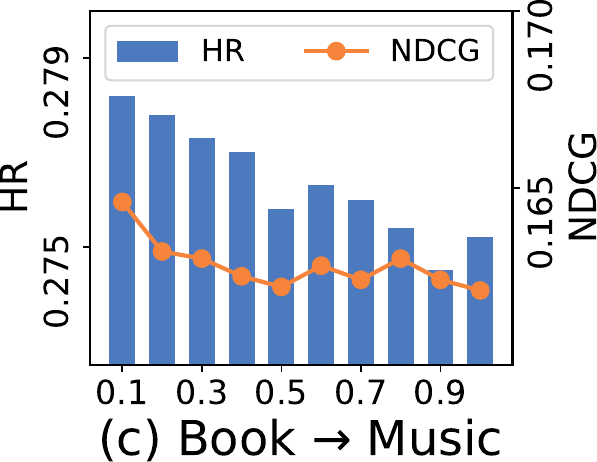}
        \end{minipage}}
        \vspace{0.5em}
        
	\caption{Effect of changing the strength ($\alpha$) of $b_{\textit{heat}}$.}
	\label{alpha}
	\end{center}
\vspace{-0.9em}
\end{figure}

\begin{figure}[t]
\setlength{\abovecaptionskip}{0.cm}
	\begin{center}
        \subfigure
        {\begin{minipage}[b]{.32\linewidth}
        \centering
        \includegraphics[scale=0.265]{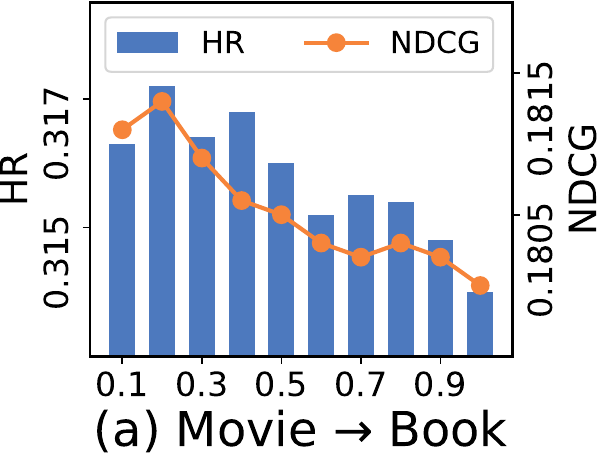}
        \end{minipage}}
        \subfigure
        {\begin{minipage}[b]{.32\linewidth}
        \centering
        \includegraphics[scale=0.265]{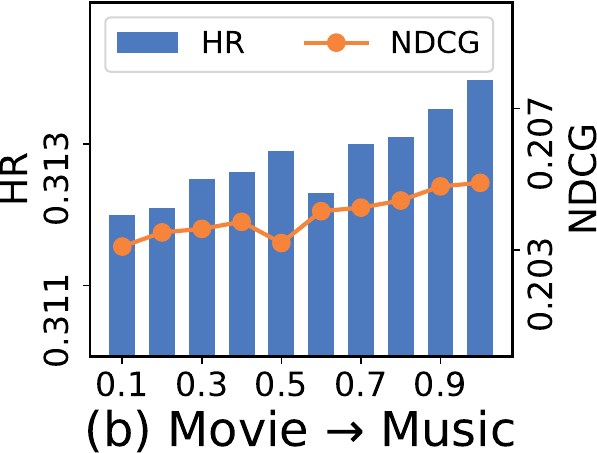}
        \end{minipage}}
        \subfigure
        {\begin{minipage}[b]{.32\linewidth}
        \centering
        \includegraphics[scale=0.265]{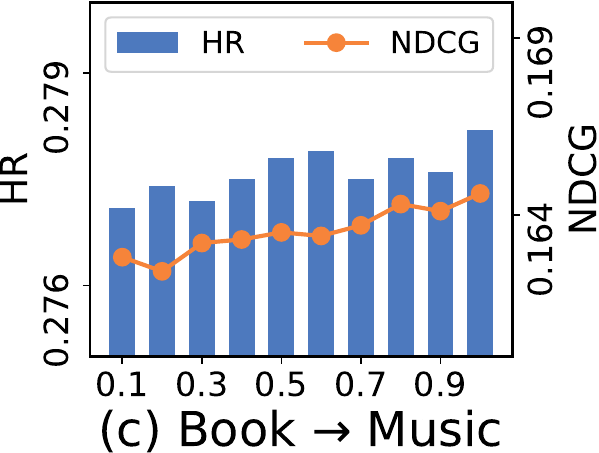}
        \end{minipage}}
        \vspace{0.5em}
        
	\caption{Effect of changing the strength ($\beta$) of $b_{\textit{ideal}}$.}
	\label{beta}
	\end{center}
\vspace{-0.9em}
\end{figure}

\begin{table*}[h]
\centering
\footnotesize
\caption{A case study of a cold-start user from \textit{Book} domain to \textit{Movie} domain in Douban.}
\setlength{\tabcolsep}{2pt}
\label{case_study}
\begin{tabular*}{0.8 \textwidth}{@{\extracolsep{\fill}}@{}c|cccccccccc@{}}
\toprule
Predictions (@10) & 1st & 2nd  & 3rd  & 4th  & 5th  &  6th &  7th  &  8th  &  9th  &  10th   \\
\midrule
\multirow{4}{*}{\textit{w}/\textit{o} $b_{\textit{ideal}}$} & \multirow{2}{*}{\tikz[baseline={(current bounding box.center)}]{\node[draw=orange, dashed, line width=0.4mm, inner sep=1.9] {\includegraphics[width=0.65cm]{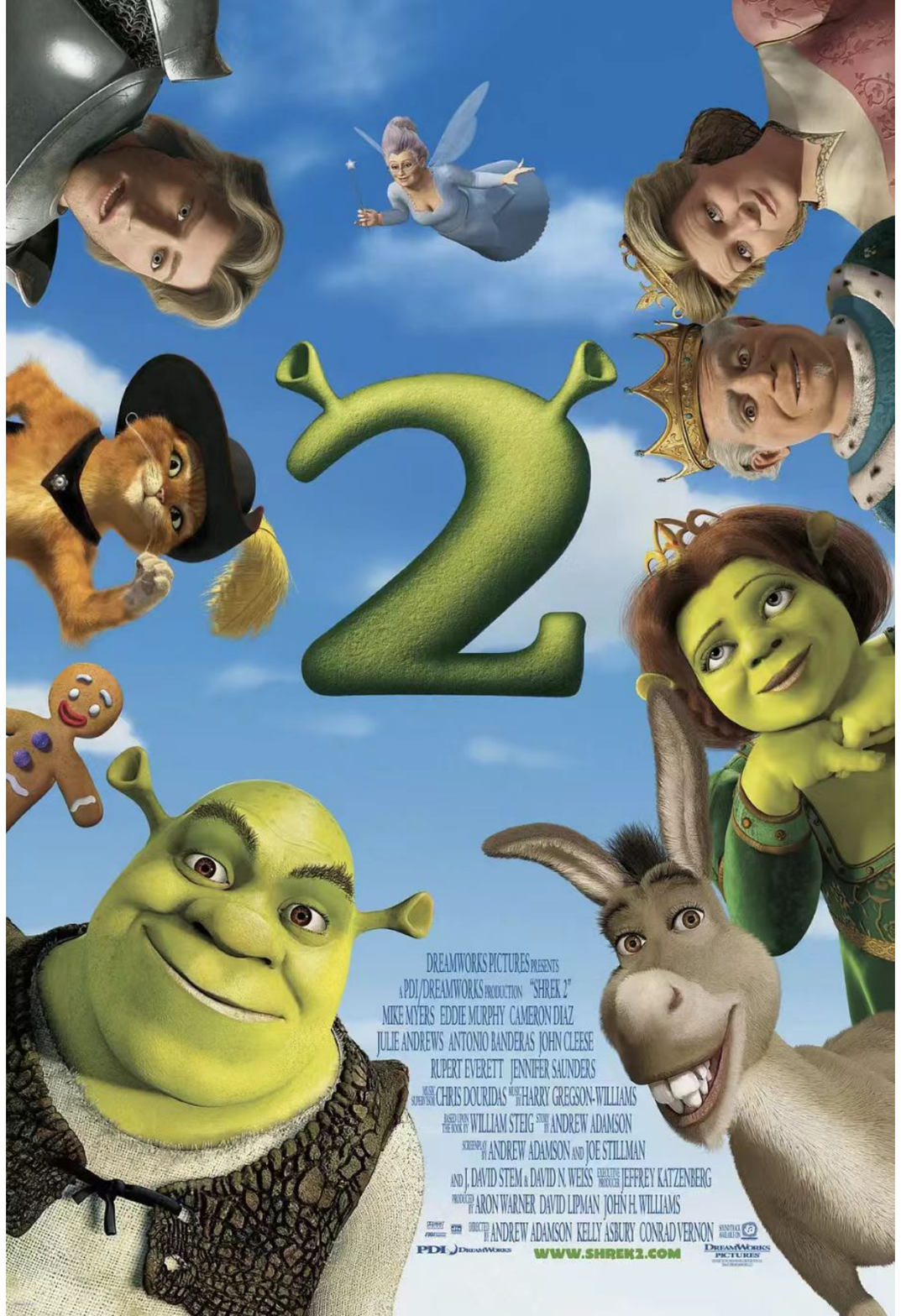}};}}  & \multirow{2}{*}{\tikz[baseline={(current bounding box.center)}]{\node[draw=blue, dashed, line width=0.4mm, inner sep=1.9] {\includegraphics[width=0.65cm]{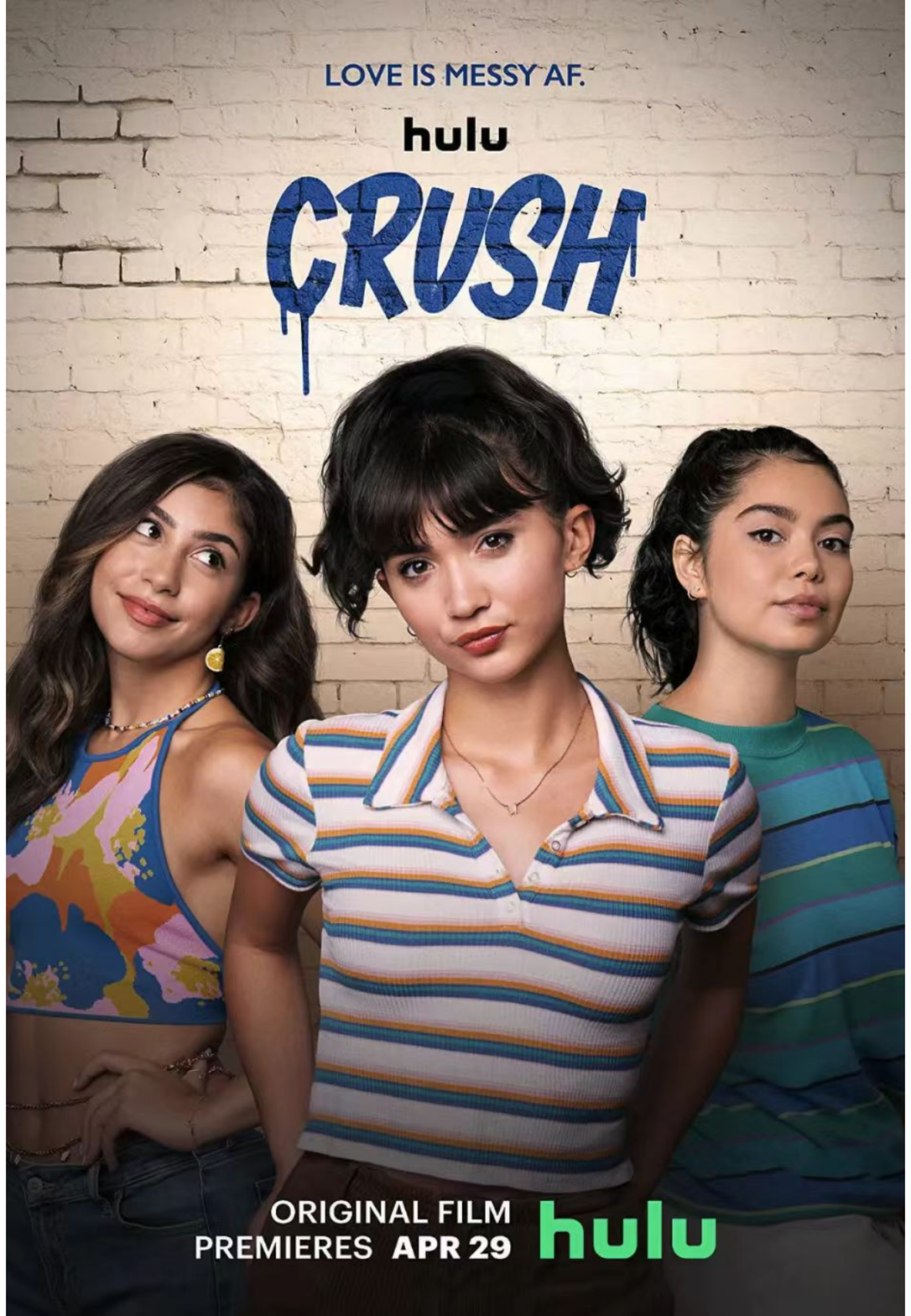}};}}  & \multirow{2}{*}{\tikz[baseline={(current bounding box.center)}]{\node[draw=blue, dashed, line width=0.4mm, inner sep=1.9] {\includegraphics[width=0.65cm]{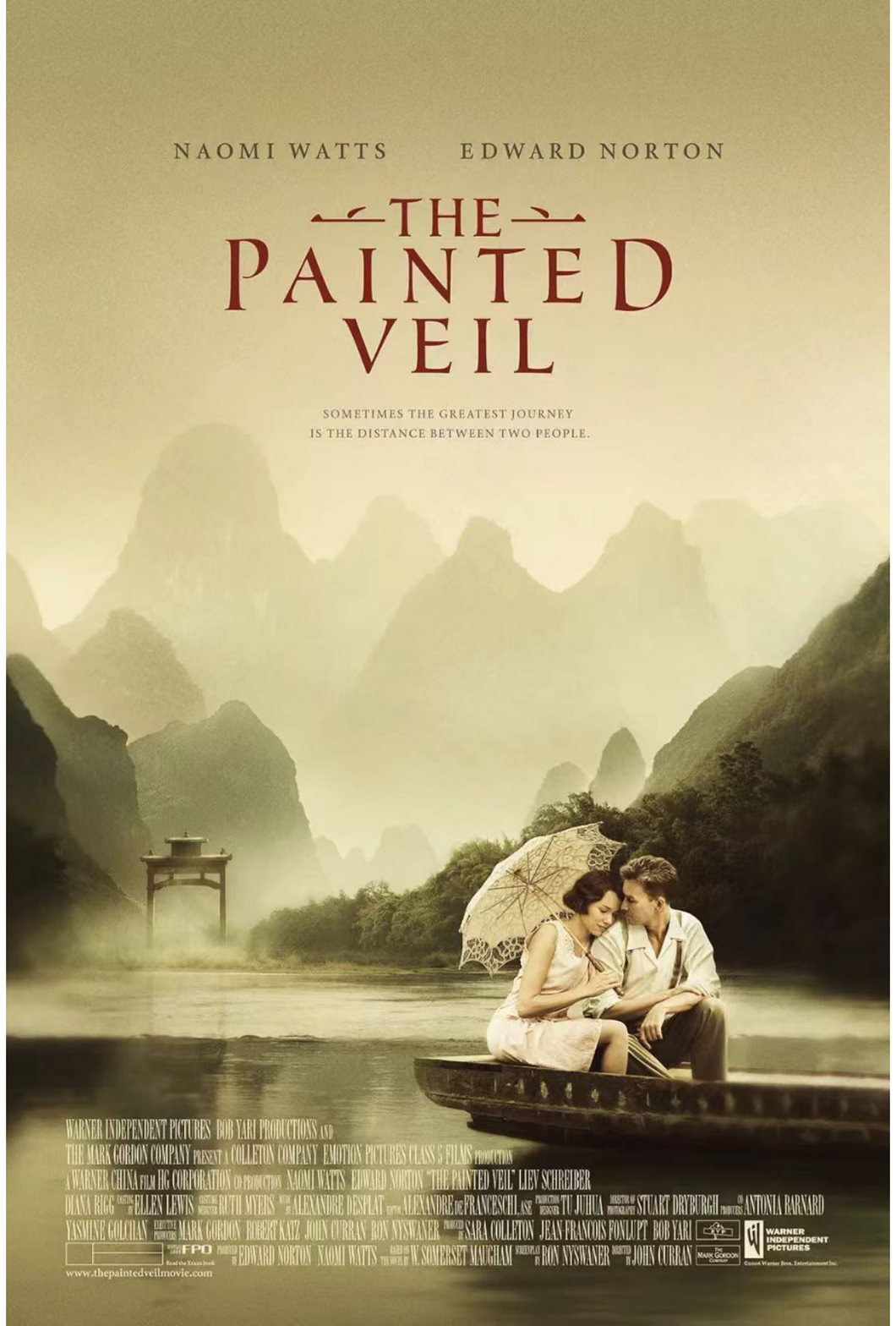}};}}  & \multirow{2}{*}{\tikz[baseline={(current bounding box.center)}]{\node[draw=red, line width=0.6mm, inner sep=1.9] {\includegraphics[width=0.65cm]{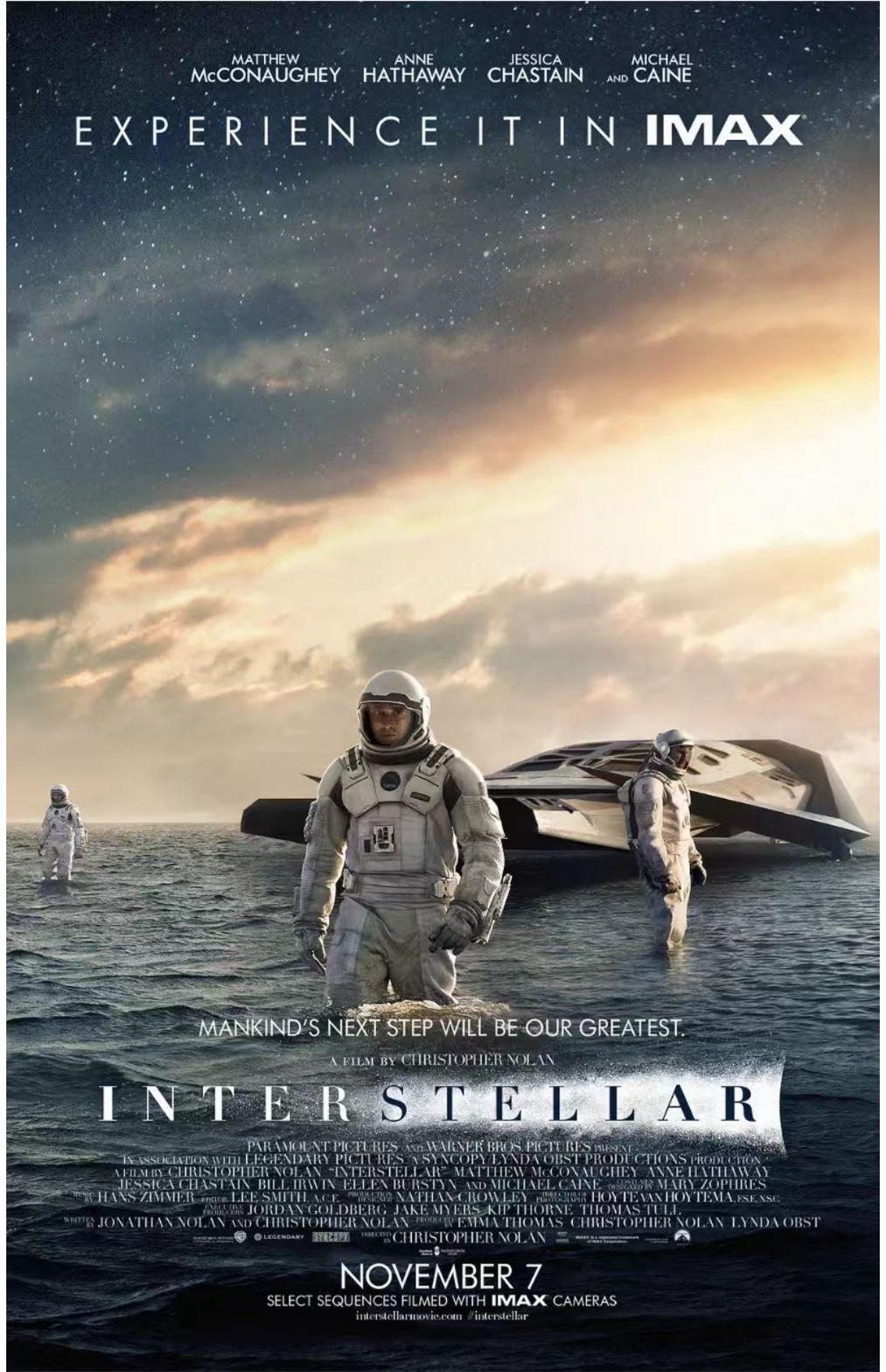}};}} & \multirow{2}{*}{\tikz[baseline={(current bounding box.center)}]{\node[draw=white, line width=0.4mm, inner sep=1.9] {\includegraphics[width=0.65cm]{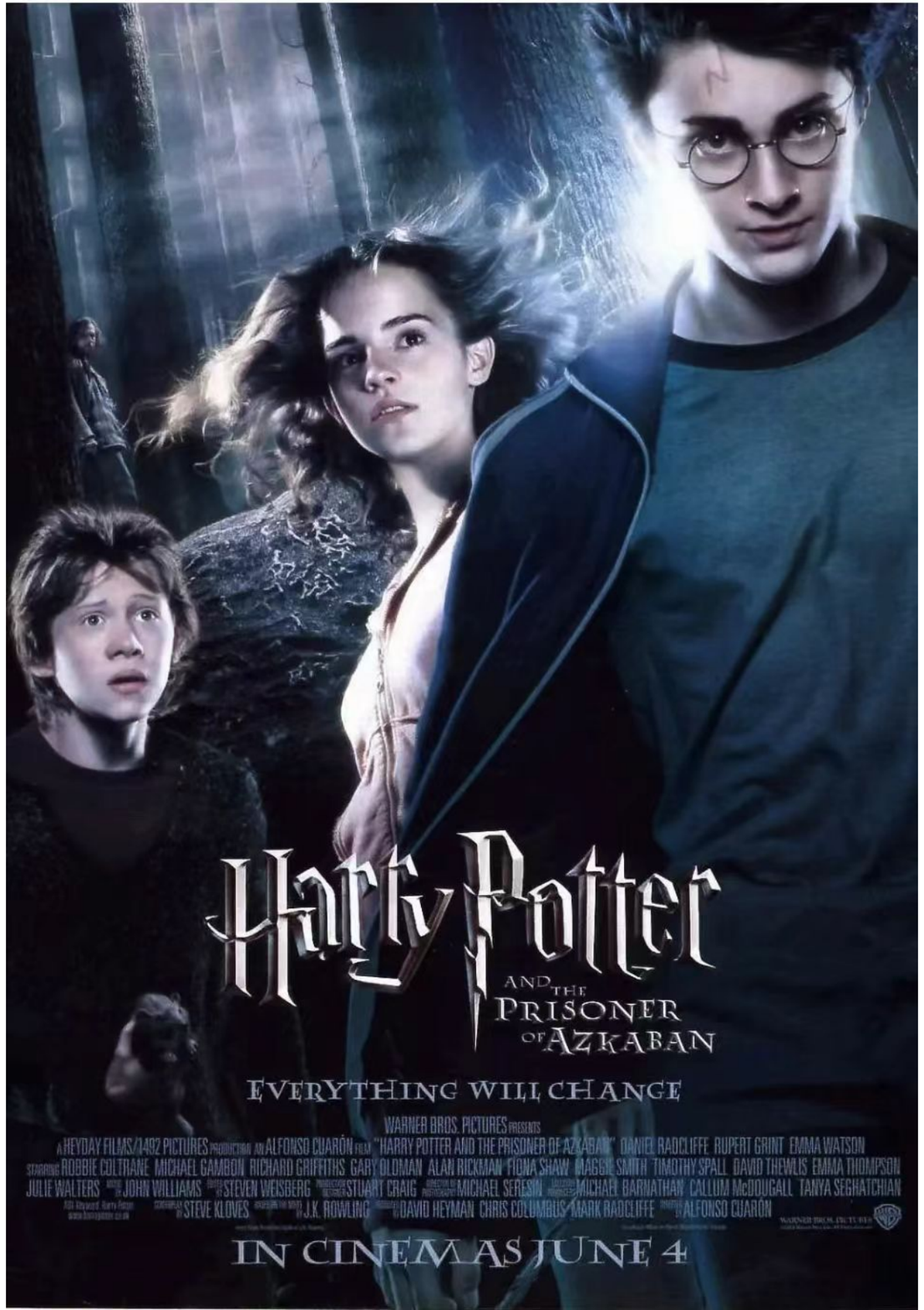}};}}  & \multirow{2}{*}{\tikz[baseline={(current bounding box.center)}]{\node[draw=white, line width=0.4mm, inner sep=1.9] {\includegraphics[width=0.65cm]{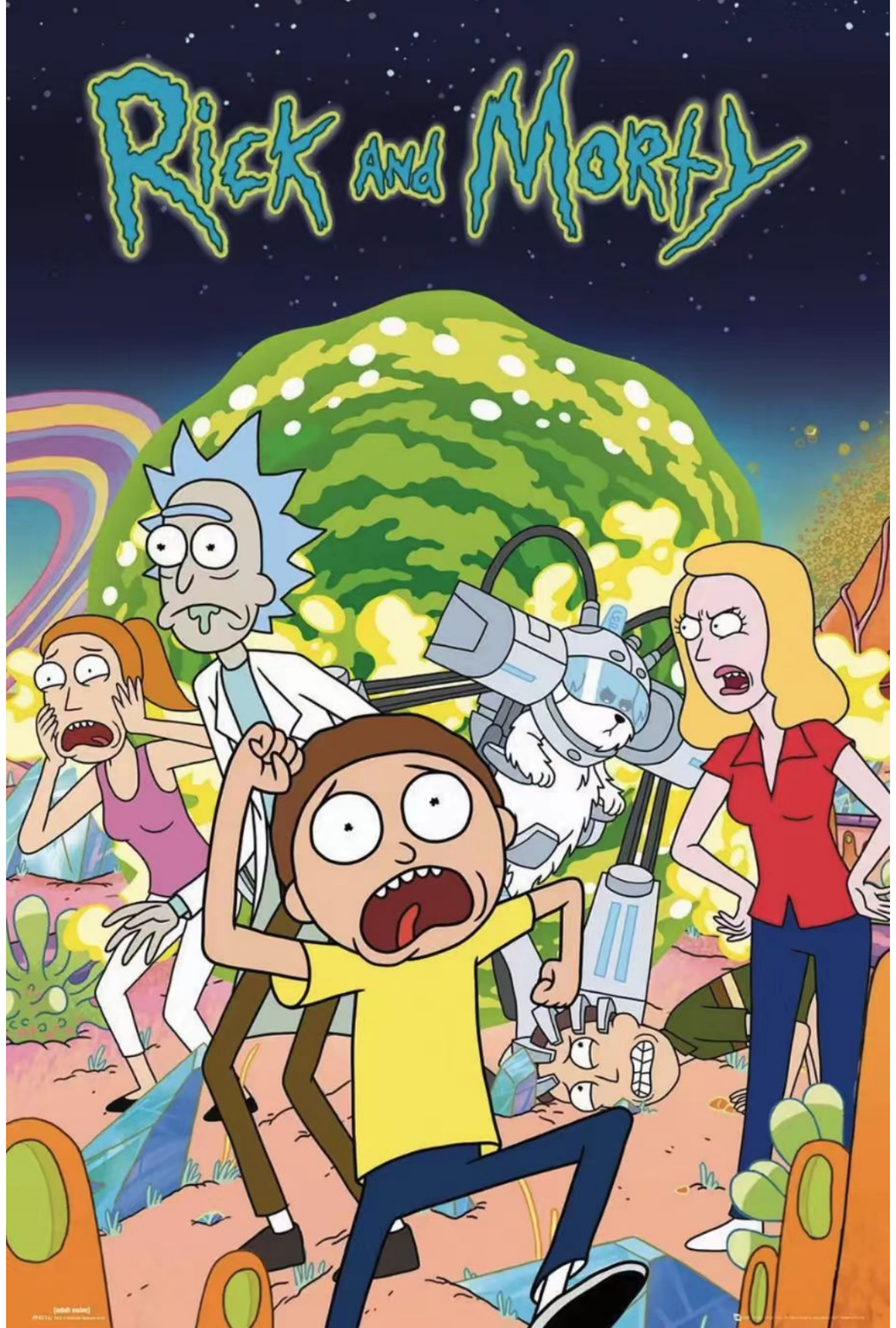}};}}  & \multirow{2}{*}{\tikz[baseline={(current bounding box.center)}]{\node[draw=white, line width=0.4mm, inner sep=1.9] {\includegraphics[width=0.65cm]{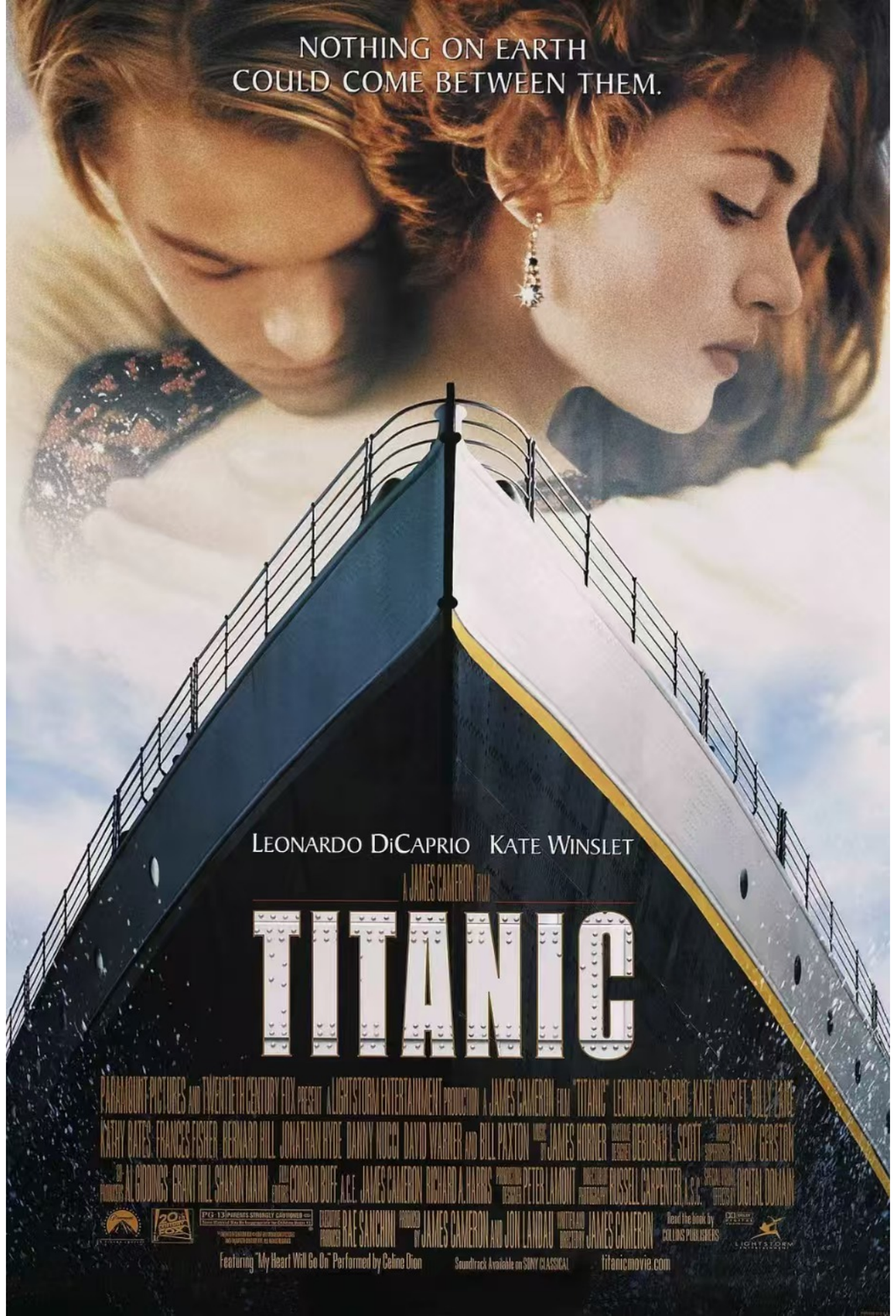}};}}  & \multirow{2}{*}{\tikz[baseline={(current bounding box.center)}]{\node[draw=white, line width=0.4mm, inner sep=1.9] {\includegraphics[width=0.65cm]{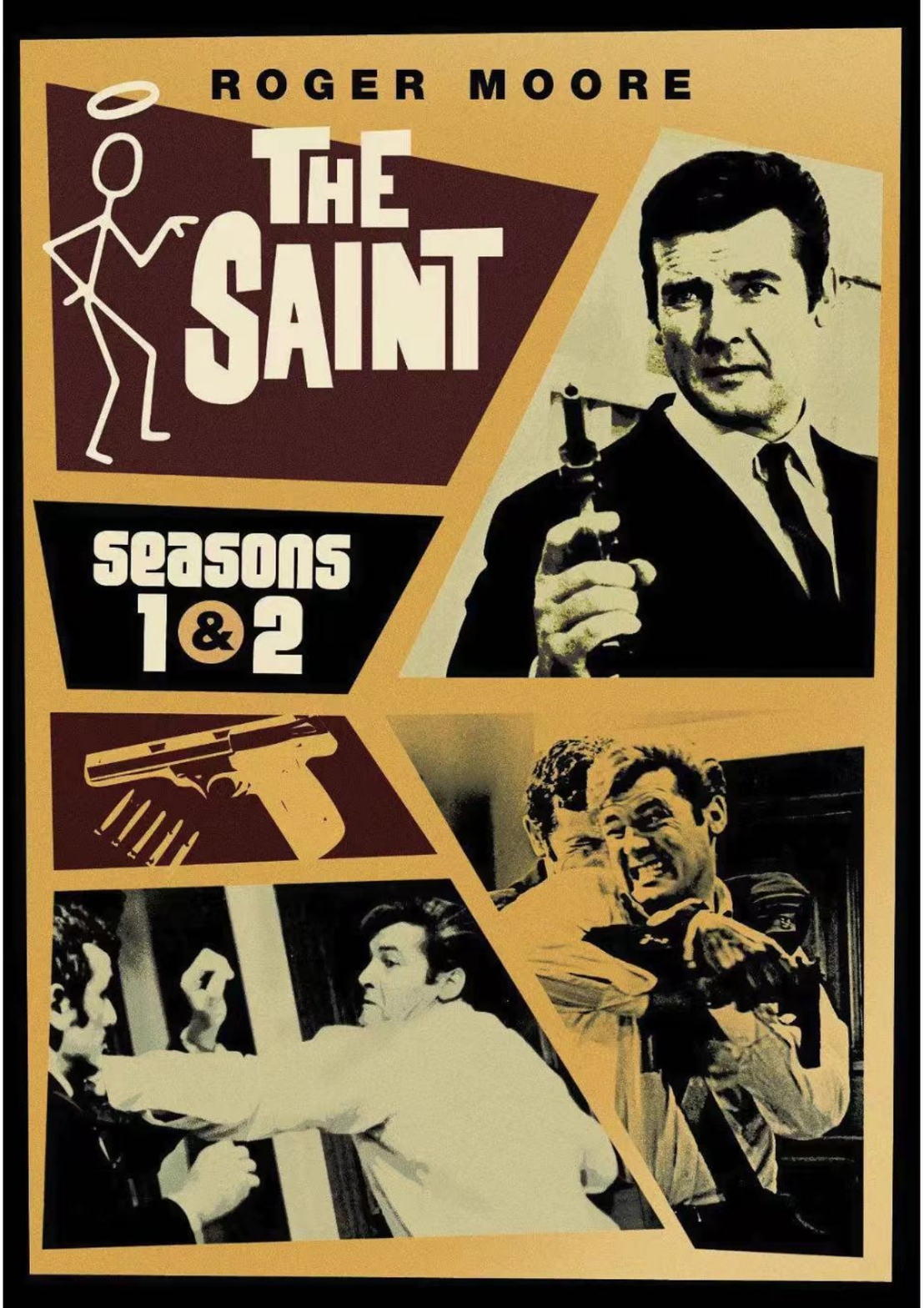}};}}  & \multirow{2}{*}{\tikz[baseline={(current bounding box.center)}]{\node[draw=white, line width=0.4mm, inner sep=1.9] {\includegraphics[width=0.65cm]{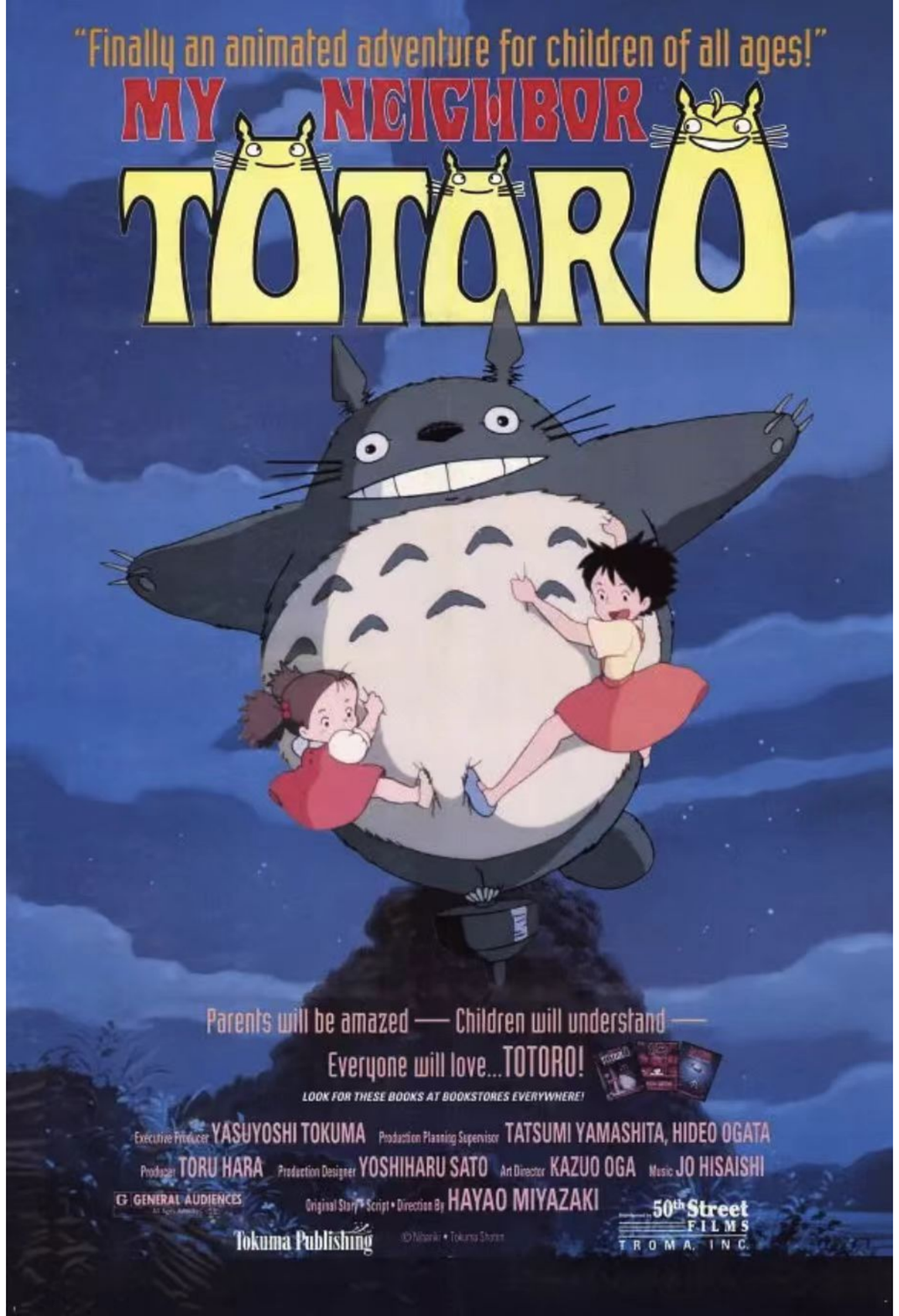}};}}  & \multirow{2}{*}{\tikz[baseline={(current bounding box.center)}]{\node[draw=white, line width=0.4mm, inner sep=1.9] {\includegraphics[width=0.65cm]{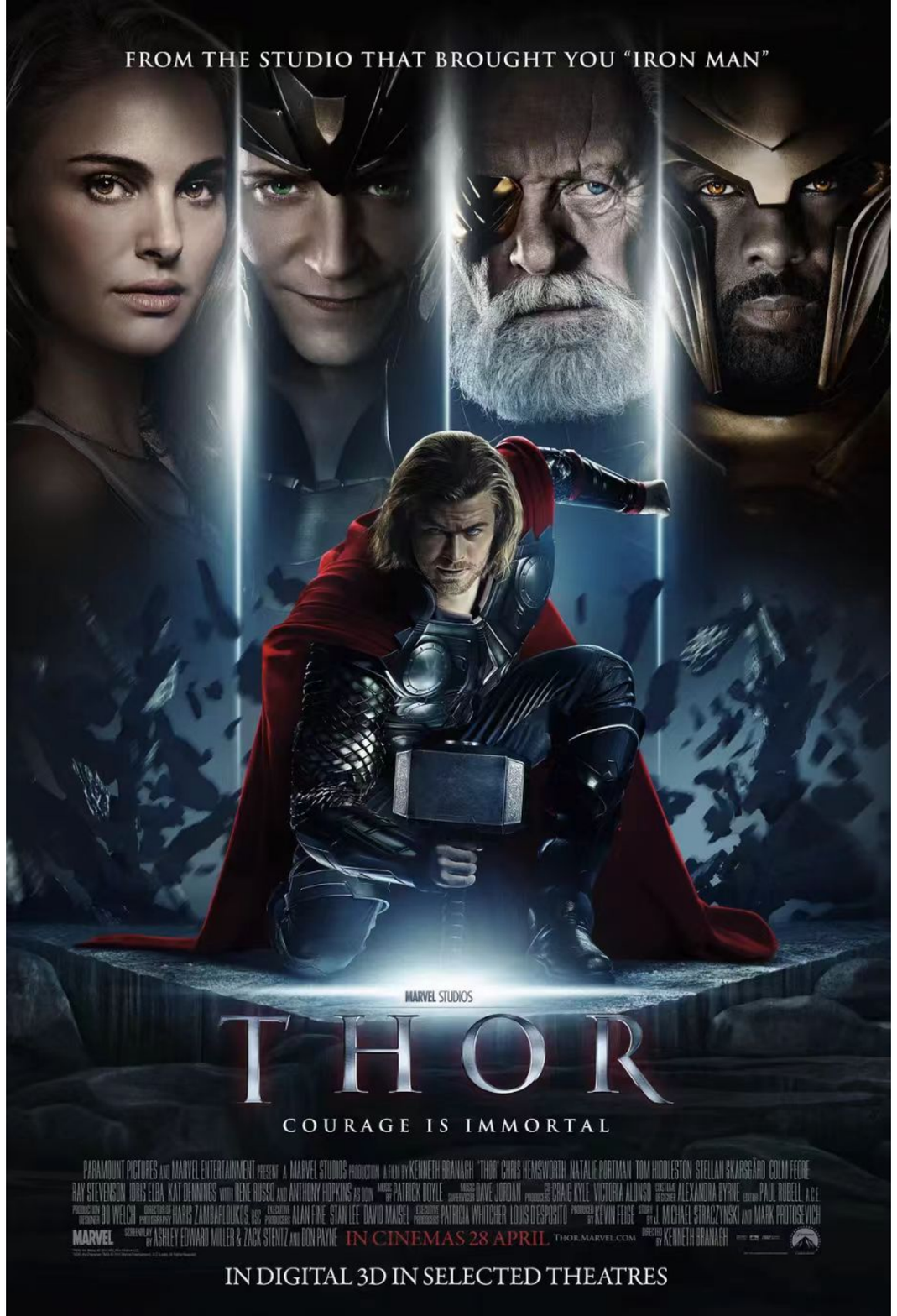}};}}   \\
 &  &    &  &    &  &      \\
 &  &   &   &    &  &     \\
 &  &    &  &    &  &      \\
\cmidrule{1-1}
\multirow{4}{*}{S$^2$CDR} & \multirow{2}{*}{\tikz[baseline={(current bounding box.center)}]{\node[draw=red, line width=0.6mm, inner sep=1.9] {\includegraphics[width=0.65cm]{figure/Adventure/Interstellar_adventure.pdf}};}}  & \multirow{2}{*}{\tikz[baseline={(current bounding box.center)}]{\node[draw=white, line width=0.4mm, inner sep=1.9] {\includegraphics[width=0.65cm]{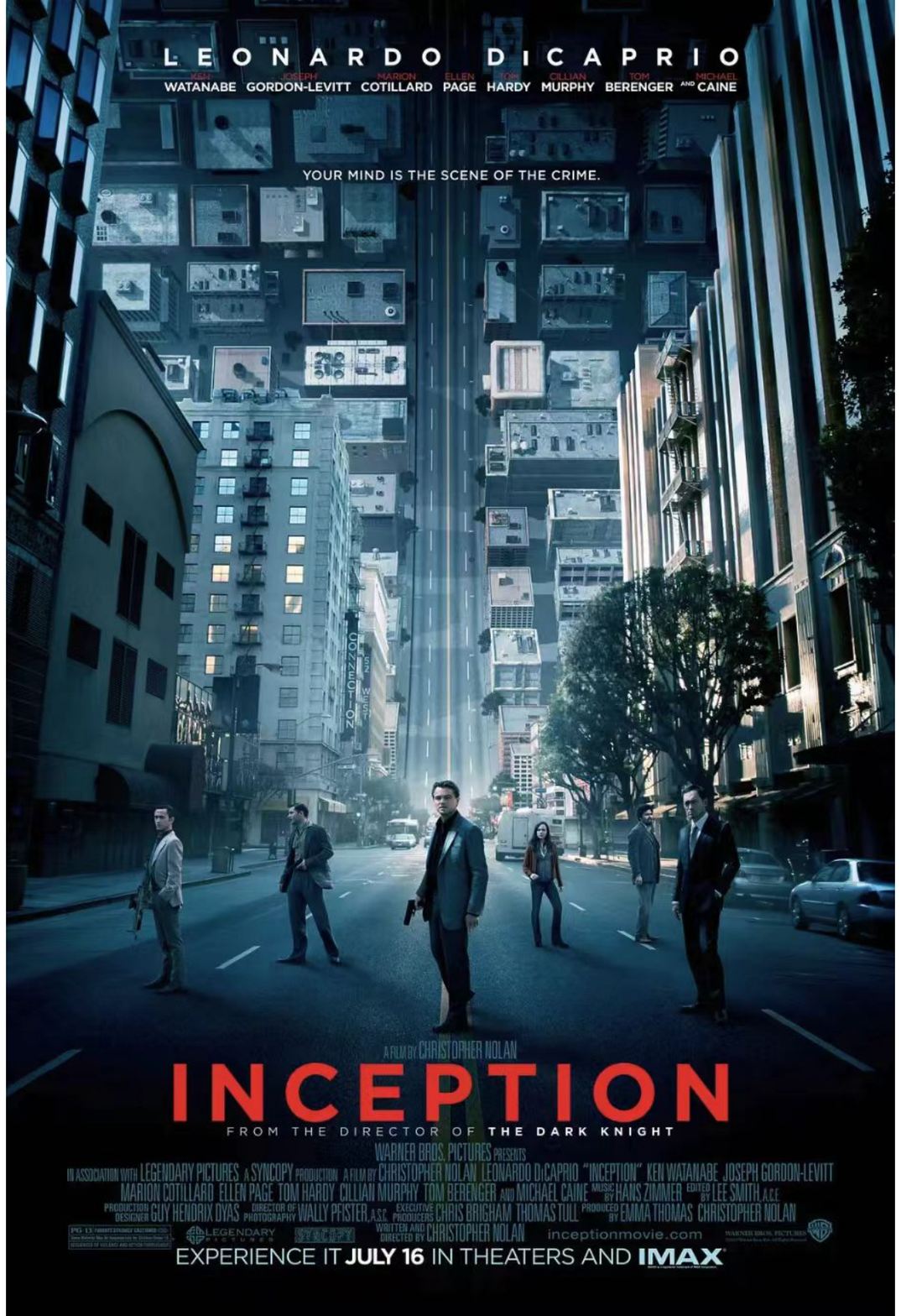}};}}  & \multirow{2}{*}{\tikz[baseline={(current bounding box.center)}]{\node[draw=white, line width=0.4mm, inner sep=1.9] {\includegraphics[width=0.65cm]{figure/Adventure/Harry_Potter_adventure.pdf}};}}  & \multirow{2}{*}{\tikz[baseline={(current bounding box.center)}]{\node[draw=blue, dashed, line width=0.4mm, inner sep=1.9] {\includegraphics[width=0.65cm]{figure/Romantic/Crush_romantic.pdf}};}} & \multirow{2}{*}{\tikz[baseline={(current bounding box.center)}]{\node[draw=white, line width=0.4mm, inner sep=1.9] {\includegraphics[width=0.65cm]{figure/Romantic/Titanic_romantic.pdf}};}}  & \multirow{2}{*}{\tikz[baseline={(current bounding box.center)}]{\node[draw=blue, dashed, line width=0.4mm, inner sep=1.9] {\includegraphics[width=0.65cm]{figure/Romantic/The_Painted_Veil_romantic.pdf}};}}  & \multirow{2}{*}{\tikz[baseline={(current bounding box.center)}]{\node[draw=orange, dashed, line width=0.4mm, inner sep=1.9] {\includegraphics[width=0.65cm]{figure/Fantasy/Shrek_fantasy.pdf}};}}  & \multirow{2}{*}{\tikz[baseline={(current bounding box.center)}]{\node[draw=white, line width=0.4mm, inner sep=1.9] {\includegraphics[width=0.65cm]{figure/Adventure/Thor_adventure.pdf}};}}  & \multirow{2}{*}{\tikz[baseline={(current bounding box.center)}]{\node[draw=white, line width=0.4mm, inner sep=1.9] {\includegraphics[width=0.65cm]{figure/Fantasy/Rick_and_Morty_fantasy.pdf}};}}  & \multirow{2}{*}{\tikz[baseline={(current bounding box.center)}]{\node[draw=white, line width=0.4mm, inner sep=1.9] {\includegraphics[width=0.65cm]{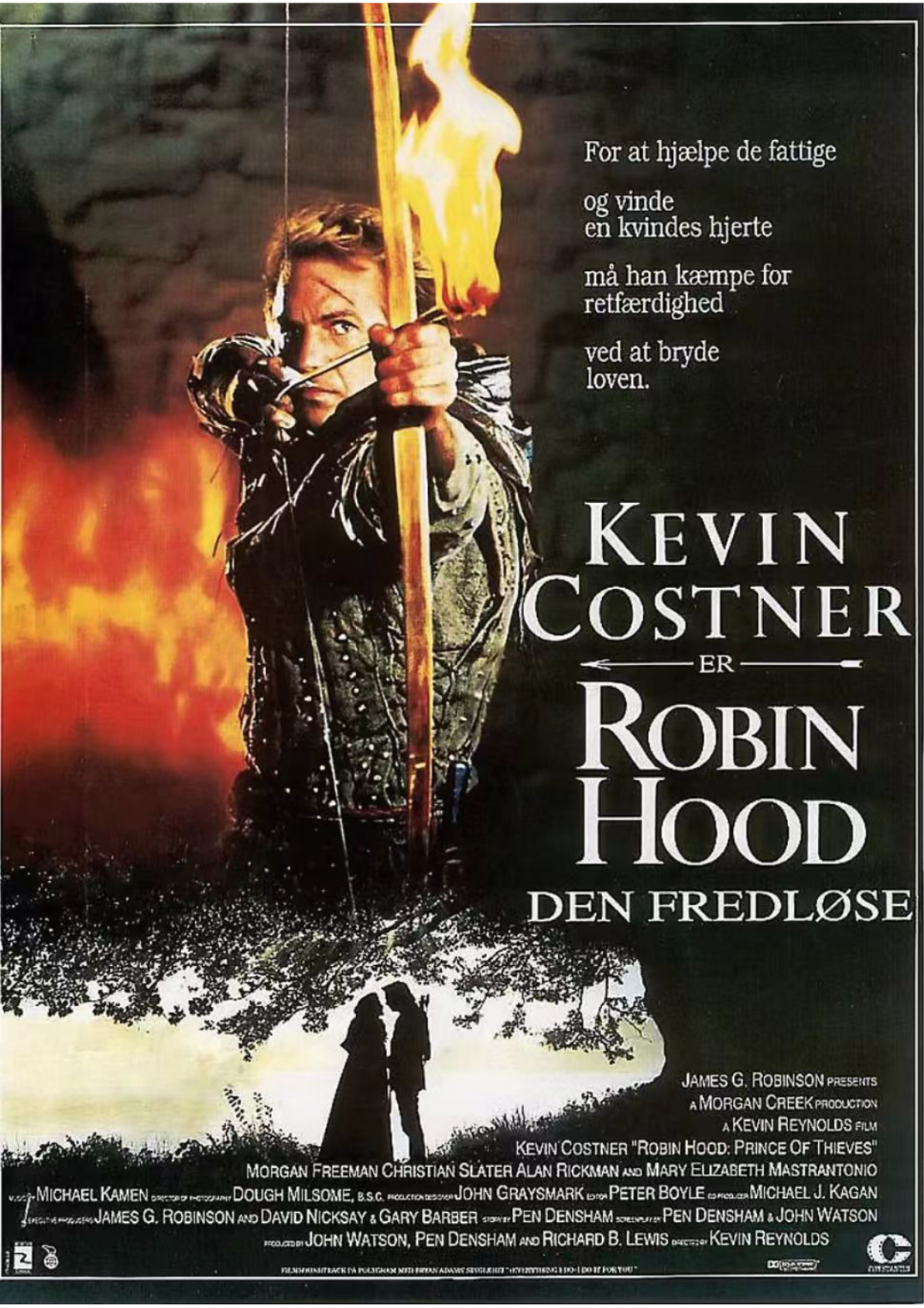}};}}       \\
 &  &    &  &    &  &      \\
 &  &    &  &    &  &      \\
 &  &    &  &    &  &      \\

\bottomrule
\end{tabular*}

\end{table*}

\begin{table*}[t]
\centering
\footnotesize
\caption{Analysis on the number of parameters and the training/inference time of DMCDR and our S$^2$CDR. Since our method does not have any training phase, we compare the pre-processing time of S$^2$CDR with the training time of DMCDR.}
\setlength{\tabcolsep}{2pt}
\label{computational_analysis}
\begin{tabular*}{0.75 \textwidth}{@{\extracolsep{\fill}}@{}c|cccccc@{}}
\toprule
\multirow{2}{*}{\color{blue}DMCDR/\color{red}S$^2$CDR} & \multicolumn{2}{c}{Movie \& Book} &  \multicolumn{2}{c}{Movie \& Music}  &  \multicolumn{2}{c}{Book \& Music} \\
& Book~$\to$~Movie & Movie~$\to$~Book  & Music~$\to$~Movie & Movie~$\to$~Music & Music~$\to$~Book & Book~$\to$~Music   \\
\midrule
\#Parameters & \multicolumn{2}{c}{\color{blue}71.52M\color{black}/\color{red}0} &  \multicolumn{2}{c}{\color{blue}21.17M\color{black}/\color{red}0}  &  \multicolumn{2}{c}{\color{blue}72.30M\color{black}/\color{red}0} \\
\#Training time & \color{blue}8.2s\color{black}/\color{red}4.3s  &   \color{blue}12.0s\color{black}/\color{red}4.8s &  \color{blue}55.9s\color{black}/\color{red}32.2s  &  \color{blue}58.7s\color{black}/\color{red}27.0s  &  \color{blue}55.3s\color{black}/\color{red}51.5s  &  \color{blue}49.6s\color{black}/\color{red}36.9s   \\
\#Inference time & \color{blue}13.1s/\color{red}11.4s &   \color{blue}11.3s/\color{red}6.7s &  \color{blue}126.4s/\color{red}109.3s &  \color{blue}115.2s/\color{red}92.6s &  \color{blue}174.8s/\color{red}150.6s &  \color{blue}168.5s/\color{red}135.2s     \\

\bottomrule
\end{tabular*}

\end{table*}

\subsection{Item-Item Similarity Analysis (RQ3)}
To better understand the ability of the heat equation $b_{\textit{heat}}$ to capture the correlations between items across domains, we construct item-item similarity graph in single domain (\textit{i.e.}, $\tilde{\bm{P}}_{\textit{source}}$ in the source domain, $\tilde{\bm{P}}_{\textit{target}}$ in the target domain) to compare with the item-item similarity graph $\tilde{\bm{P}}$ across domains (refer to Section~\ref{problem_setting}) in Table~\ref{item_item}:
\begin{itemize}
    \item \textit{w}/ $\tilde{\bm{P}}_{\textit{source}}$: This variant replaces $\tilde{\bm{P}}$ in Eq.~(\ref{heat}) and (\ref{sharpen_function}) with $\tilde{\bm{P}}_{\textit{source}}$.
    \item \textit{w}/ $\tilde{\bm{P}}_{\textit{target}}$: This variant replaces $\tilde{\bm{P}}$ in Eq.~(\ref{heat}) and (\ref{sharpen_function}) with $\tilde{\bm{P}}_{\textit{target}}$.
\end{itemize}
From the results shown in Table~\ref{item_item}, we find that S$^2$CDR consistently exhibits the best performance compared with \textit{w}/ $\tilde{\bm{P}}_{\textit{source}}$ variant and \textit{w}/ $\tilde{\bm{P}}_{\textit{target}}$ variant in all three CDR scenarios, which demonstrate the effectiveness of the $\tilde{\bm{P}}$ we designed to capture the correlations between items across domains. Another observation is that for cold-start users in the target domain, such as \textit{Music} domain in Scenario 2, $\tilde{\bm{P}}_{\textit{source}}$ variant outperforms $\tilde{\bm{P}}_{\textit{target}}$ variant by a large margin, suggesting that transfer item-item similarity across domains can benefit preference modeling for cold-start users.

We further explore how changing the strength (\textit{i.e.}, $\alpha$ in Eq.(\ref{smoothing_function})) of heat equation $b_{\textit{heat}}$ affects the model's ability to capture the correlations between items across domains in Figure~\ref{alpha}. For Movie $\to$ Book, we find the performance of S$^2$CDR improves as the value of $\alpha$ increases. The reason might be that the correlations between items in CDR scenario with fewer interactions is limited, which should be captured by a large $\alpha$. In contrast, for the other two CDR scenarios with richer interactions, a smaller $\alpha$ is enough to capture the correlations between items across domains.

\subsection{Noise Filtering Analysis (RQ4)}
To verify the effectiveness of the ideal low-pass filter $b_{\textit{ideal}}$ to filter out the high-frequency noise information, we change the strength (\textit{i.e.}, $\beta$ in Eq.~(\ref{smoothing_function})) of $b_{\textit{ideal}}$ in Figure~\ref{beta}. 
We find that a smaller $\beta$ can achieve good performance in Movie $\to$ Book. This observation can be attributed to the CDR scenarios with fewer interactions has less noise information, which can be filtered out by a smaller $\beta$. Another observation is that, since the CDR scenarios (\textit{e.g.}, Movie $\to$ Music and Book $\to$ Music) with richer interactions have more noise information, which should be filtered out by a larger $\beta$. 

We also conduct a case study to compare the noise filtering ability of \textit{w}/\textit{o} $b_{\textit{ideal}}$ variant and S$^2$CDR in Table~\ref{case_study}. Specifically, we randomly select a user from \textit{Book} domain to \textit{Movie} domain in Douban. The interaction history of this user is given in the dataset. Fortunately, the \textit{Book} domain in Douban dataset provides book label, which makes it easy for us to obtain user preference in \textit{Book} domain. Thus, we observe that the user's preference in \textit{Book} domain can be roughly divided into three types, including Adventure (five books), Fantasy (one book) and Romantic (one book), represented by red, orange and blue respectively. We also find that this user only watches adventure movies in \textit{Movie} domain, which means that the user's fantasy and romantic preferences in \textit{Book} domain are noise information. We treat this user as a cold-start user to predict potential interactions in \textit{Movie} domain, where the ground truth is indicated by a red box. From the results shown in Table~\ref{case_study}, we find that \textit{w}/\textit{o} $b_{\textit{ideal}}$ variant can not discriminate noise preferences (\textit{i.e.}, Fantasy preference in orange box and Romantic preference in blue box). In contrast, our S$^2$CDR successfully filters out the noise information, further demonstrating the effectiveness of $b_{\textit{ideal}}$.

\subsection{Computational Efficiency Analysis (RQ5)}
We conduct a comprehensive evaluation of the computational efficiency for our S$^2$CDR and the DMs-based methods DMCDR, including the trainable parameters, training time and inference time in Table~\ref{computational_analysis}. 
As a non-parametric method, there's nothing to train in our S$^2$CDR. In contrast, DMCDR requires high trainable parameters as shown in Table~\ref{computational_analysis}. In our method, we need a pre-processing step to calculate $\bm{R}^{st}$, $\tilde{\bm{P}}$ and so on, thus we use the pre-processing time to compare with the training time of DMCDR. From the results shown in Table~\ref{computational_analysis}, we also observe that the training and inference time costs of our method are lower than DMCDR, demonstrating the high computational efficiency of S$^2$CDR.

\section{Related Work}

Traditional \textbf{cross-domain recommendation} (CDR) aims to tackle the data sparsity problem~\cite{disencdr,conet,tdar} for overlapping users in the target domain. These CDR methods utilize the auxiliary information in the source domain to improve the recommendation performance in the sparse target domain. However, they cannot make recommendation for cold-start users with no interactions in the target domain.

As a more challenging problem, using cross-domain recommendation to solve the cold-start issue~\cite{emcdr,sscdr,ptupcdr,udmcf,cdrnp,dmcdr,nf-npcdr} has attracted the attention of many researchers. Classical CDR works on this can be roughly divided into two types, \textit{i.e.}, mapping-based methods and meta-based methods. Mapping-based methods~\cite{emcdr,sscdr,udmcf} aims to align and transform the user/item representations across domains. Specifically, EMCDR~\cite{emcdr} is the most popular mapping-based method, which first pre-trains user/item representations in both domains, and then learns a mapping function to transfer user preference across domains. SSCDR~\cite{sscdr} further extends EMCDR by mapping user/item representations with deep metric learning in a semi-supervised learning. UDMCF~\cite{udmcf} proposes unbalance distribution optimal transport with typical subgroup discovering algorithm for user distribution mapping. Meta-based methods~\cite{ptupcdr,tmcdr,remit,cdrnp} adopt meta-learning paradigm to map features across domains. TMCDR~\cite{tmcdr} and PTUPCDR~\cite{ptupcdr} propose mapping functions for each user to transfer their personalized preference with meta networks. 
CDRNP~\cite{cdrnp} introduces neural process to capture the preference correlations among the overlapping and cold-start users. In addition to the above two paradigms, DMCDR~\cite{dmcdr} adopts diffusion models to generate personalized user representation in the target domain under the guidance of user preference in the source domain.

\textbf{GSP-based collaborative filtering} has attracted significant attention in recommendation systems~\cite{gf-cf,pgsp,bspm,higsp,fagsp,cgsp,fire,giffcf} due to its outstanding performance and high computational efficiency. Specifically, GF-CF~\cite{gf-cf} proposes a unified graph convolution-based framework and integrates a linear filter and an ideal low-pass filter for recommendation. PGSP~\cite{pgsp} adopts a mixed-frequency graph filter to smooth signals globally and locally for user preference modeling. FIRE~\cite{fire} improves the computational efficiency and captures the temporal dynamics of user/item by designing temporal information and side information filters. 
BSPM~\cite{bspm} introduces a blurring-sharpening process to improve recommendation performance. CGSP~\cite{cgsp} employs a cross-domain similarity graph to mitigate the intrinsic discrepancy between source and target domains. 
FaGSP~\cite{fagsp} designs two filter modules to capture both unique and common user/item features and high-order neighborhood information. HiGSP~\cite{higsp} recognize user-matched interaction patterns with the linearly combining of cluster-wise and globally-aware filters.


Different from prior methods, we propose the smoothing-sharpening process model, a novel training-free framework which innovatively adapts the graph signal processing (GSP) theory into CDR to capture the item correlation and intrinsic user preference for transfer.
Although no training is required, S$^2$CDR still achieves SOTA, confirming the efficiency and effectiveness of corruption-recovery paradigm supported by GSP and inspiring a new perspective.

\section{Conclusion and Future Work}
This paper proposes a novel smoothing-sharpening process model (S$^2$CDR) for CDR to cold-start users, which applies a corruption-recovery architecture and can be solved by the ODEs in a continuous manner. The smoothing process gradually corrupts the user-item/item-item interaction matrices derived from both domains into smoothed preference signals in a noise-free manner with tailor-designed graph filters, 
and the sharpening process iteratively sharpens the preference signals to recover potential interactions for cold-start users.
As a non-parametric method, our S$^2$CDR has significant advantages in both effectiveness and efficiency.
In the future, we will explore the smoothing and sharpening process for more challenging tasks, \textit{e.g.}, multi-domain recommendation.

\section*{Acknowledgement}
This work was supported by the National Natural Science Foundation of China (No.62406319).

\balance
\bibliographystyle{ACM-Reference-Format}
\bibliography{sample-base-extend.bib}

\clearpage

\appendix

\section{Appendix}

\subsection{Diffusion Models}\label{diffusion_models}
As a powerful family of generative models, diffusion models (DMs) have shown great performance in the field of RS~\cite{dreamrec,drm,mcdrec,dmcdr}, which include the forward and reverse processes.

\subsubsection{\textbf{Forward Process}}
Given the input sample $\bm{x}_0 \sim q(\bm{x}_0)$, the forward process corrupts $\bm{x}_0$ by adding Gaussian noise:
\begin{equation}
\begin{split}
q(\bm{x}_t|\bm{x}_{t-1}) &= \mathcal{N}(\bm{x}_t;\sqrt{1-\beta_t}\bm{x}_{t-1},\beta_t\textbf{I}),
\end{split}
\label{}
\end{equation}
where $t\in\{1,\dots,T\}$ denotes the diffusion steps, $\beta_t\in(0,1)$ is the noise scale in step $t$, and $\mathcal{N}$ is the Gaussian distribution.

\subsubsection{\textbf{Reverse Process}}
Starting from the noise data $\bm{x}_T$, the reverse process iteratively eliminates the noise to reconstruct $\bm{x}_0$:
\begin{equation}
\begin{split}
p_{\theta}(\bm{x}_{t-1}|\bm{x}_t) &= \mathcal{N}(\bm{x}_{t-1};\bm{\mu}_\theta(\bm{x}_t,t),\textstyle\sum_\theta(\bm{x}_t,t)),
\end{split}
\label{}
\end{equation}
where $\bm{\mu}_\theta(\bm{x}_t,t)$ and $\sum_\theta(\bm{x}_t,t)$ are the mean and variance of the Gaussian distribution parameterized by a neural network.

DMCDR~\cite{dmcdr} is the representative work that applies DMs to solve the cold-start problem in CDR. During inference stage, DMCDR can generate personalized user representation from noise data using only the reverse process. 
Since DMCDR and our S$^2$CDR have similar processing mechanism, we formally compare them as follows:
\begin{itemize}
    \item Both models employ a series of corruption-recovery processes. However, DMCDR deal with the user/item embeddings, which requires a time-consuming training phase. In contrast, since our S$^2$CDR directly handles the user-item/item-item interaction matrix, there is nothing to learn and it is computationally efficient. 
    \item Both models expect to discover personalized user preference during the reverse process. Specifically, the reverse process of DMCDR is a generative process, which can generate personalized representation for cold-start users. In our S$^2$CDR, unknown user-item interactions are predicted during the sharpening process.
\end{itemize}

\begin{table}[t]
\centering
\normalsize
\caption{Notations.}
\setlength{\tabcolsep}{2pt}
\label{appendix_notations}
\begin{tabular*}{0.45 \textwidth}{@{\extracolsep{\fill}}@{}cl@{}}
\toprule
\textbf{Notation} & \textbf{Description}\\
\midrule
$\mathcal{D}^s$, $\mathcal{D}^t$  & Interaction data \\
$\mathcal{U}^s$, $\mathcal{U}^t$ &  User set \\
$\mathcal{V}^s$, $\mathcal{V}^t$  &  Item set \\
$\mathcal{Y}^s$, $\mathcal{Y}^t$ &  Rating scores \\
$\bm{R}^s$, $\bm{R}^t$  &  User-item interaction matrix \\
$\mathcal{U}^o$  &  Overlapping users \\
$\mathcal{U}^{s\setminus o}$, $\mathcal{U}^{t\setminus o}$  &  Cold-start users \\
$\bm{R}^{st}$, $\tilde{\bm{R}}^{st}$  &  Unified user-item interaction matrix  \\
$\tilde{\bm{P}}$  &  Normalized item-item adjacency matrix  \\
$\hat{\bm{R}}^{st}$  &  Inferred interaction matrix  \\
\midrule
$b_{\textit{heat}}$  &  Heat equation \\
$b_{\textit{ideal}}$  &  Ideal low-pass filter \\
$b_{\textit{smooth}}$, $b_{\textit{sharpen}}$  & Smoothing and sharpening functions \\
$\alpha$, $\beta$  &  Ration of $b_{\textit{heat}}$ and $b_{\textit{ideal}}$  \\
$T_b$, $T_h$  &  Terminal time  \\
${T_b}/s$, ${T_h}/s$  &  Number of steps  \\

\bottomrule
\end{tabular*}
\end{table}

\subsection{Notations}\label{notations}
We summarize the notations used in this paper in Table~\ref{appendix_notations}.


\begin{algorithm}[tb]
\caption{The Overall Procedure of S$^2$CDR}
\label{appendix_algorism}
\textbf{Input}: Overlapping users $\mathcal{U}^o$; Cold-start users $\mathcal{U}^{s\setminus o}$, $\mathcal{U}^{t\setminus o}$; Items set $\mathcal{V}^s$, $\mathcal{V}^t$; User-item interaction matrix $\bm{R}^s$, $\bm{R}^t$. \\
\textbf{Output}: Inferred interaction matrix $\hat{\bm{R}}^{st}$.\quad\quad\quad\quad\quad\quad\quad\quad\quad\quad\quad\quad\\
\begin{algorithmic}[1] 
\STATE Construct matrix $\bm{R}^{st}$ and $\tilde{\bm{P}}$. \\
\STATE Perform $b_{\textit{heat}}$ and $b_{\textit{ideal}}$ in Eq.~(\ref{heat}) and Eq.~(\ref{ideal}). \\ 
\STATE Apply ODE solvers to solve Eq.~(\ref{smoothing_process_final}) for $\frac{T_b}{s}$ steps.\\
\STATE Perform $b_{\textit{sharpen}}$ in Eq.~(\ref{sharpen_function}).\\
\STATE Apply ODE solvers to solve Eq.~(\ref{sharpening_process}) for $\frac{T_h}{s}$ steps.\\
\STATE Derive the inferred matrix $\hat{\bm{R}}^{st}$.
\end{algorithmic}
\end{algorithm}

\subsection{Training Algorism}\label{alg1}
We give the overall procedure of S$^2$CDR in Algorism~\ref{appendix_algorism}.

\begin{table}[t]
\centering
\footnotesize
\caption{Statistics of three CDR scenarios (\textit{Overlap} denotes the number of overlapping users).}
\setlength{\tabcolsep}{4.5pt}
{
\begin{tabular*}{0.48 \textwidth}{@{\extracolsep{\fill}}@{}cccccc@{}}
\toprule
{\bf Scenarios}  & \bf Domain  &  {\bf Users}  &  {\bf Overlap}  &  {\bf Items}  &  {\bf Ratings}\\
\midrule
\multirow{2}{*}{\textbf{Douban:~}\textit{\textbf{Movie}}~$\&$~\textit{\textbf{Book}}}  & Movie  &2,712   &\multirow{2}{*}{2,209}   &{34,893}  &1,278,401\\ 
&Book   &2,212  &  &  {95,872}  &227,251\\
\midrule
\multirow{2}{*}{\textbf{Amazon:~}\textit{\textbf{Movie}}~$\&$~\textit{\textbf{Music}}}  & Movie  &123,960   &\multirow{2}{*}{18,031}   & 
 {50,052}  &1,697,533\\ 
& Music  &75,258  &  &  {64,443}  &1,097,592\\ 
\midrule
\multirow{2}{*}{\textbf{Amazon:~}\textit{\textbf{Book}}~$\&$~\textit{\textbf{Music}}}  & Book  &603,668   &\multirow{2}{*}{16,738}   &{367,982}  &8,898,041\\ 
&Music   &75,258  &  &  {64,443}  &1,097,592\\ 
\bottomrule
\end{tabular*}
}
\label{appendix_dataset}
\end{table}

\subsection{Details of three CDR scenarios}\label{dataset}
We summarize the details of three CDR scenarios in Table~\ref{appendix_dataset}.

\subsection{Details of Baselines}\label{baselines}
The details of the sate-of-the-art baselines used in this paper are summarized as follows:
(1) \textbf{NeuMF}~\cite{ncf} is a popular single-domain recommendation method that leverages neural networks to tackle the collaborative filtering problem. (2) \textbf{EMCDR}~\cite{emcdr} is the first mapping-based CDR method, which utilizes a mapping function to transfer features across domains. (3) \textbf{DCDCSR}~\cite{dcdcsr} employs the DNN guided by sparsity degrees of individual users and items to map the latent factors across domains or systems. (4) \textbf{SSCDR}~\cite{sscdr} adopts a semi-supervised approach to learn the cross-domain relationship of users and items. (5) \textbf{TMCDR}~\cite{tmcdr} designs a transfer-meta framework with a pre-training stage and a meta stage to implicitly transform the user embedding. (6) \textbf{LACDR}~\cite{lacdr} proposes a low-dimensional alignment strategy to align the overlapping users representations in low-dimensional space. (7) \textbf{DOML}~\cite{doml} adopts dual learning and metric learning for knowledge transfer in an iterative manner. (8) \textbf{BiTGCF}~\cite{bitgcf} utilizes the graph neural networks to fuse users common and domain-specific preference with feature transfer layer. (9) \textbf{PTUPCDR}~\cite{ptupcdr} follows meta-based paradigm, which designs mapping functions for each user to transfer their personalized preference. (10) \textbf{CDRIB}~\cite{cdrib}
leverages the information bottleneck principle to derive the unbiased user representations. (11) \textbf{UDMCF}~\cite{udmcf} combines the collaborative filtering and distribution mapping process to solve the cold-start problem. (12) \textbf{GF-CF}~\cite{gf-cf} is the first method that apply graph signal processing for GCN-based collaborative filtering. (13) \textbf{PGSP}~\cite{pgsp} introduces a mixed-frequency graph filter to capture user preference in the high-frequency components. (14) \textbf{DMCDR}~\cite{dmcdr} is the state-of-the-art method for CDR which employs diffusion models to generate user representation in the target domain under the guidance of user preference in the source domain.

\subsection{Hyper-parameters}\label{best_hyperparameters}
The best hyper-parameter configurations in each CDR scenario are as follows: In $\textit{Movie}~$\&$~\textit{Book}$, $\alpha=0.3$, $\beta=0.1$, $T_b=1$, $\frac{T_b}{s}=2$, $T_h=2.4$, and $\frac{T_h}{s}=1$ in the Movie domain, $\alpha=0.9$, $\beta=0.2$, $T_b=1$, $\frac{T_b}{s}=1$, $T_h=2.5$, and $\frac{T_h}{s}=1$ in the Book domain. In $\textit{Movie}~$\&$~\textit{Music}$, $\alpha=0.3$, $\beta=1$, $T_b=1$, $\frac{T_b}{s}=1$, $T_h=2.8$, and $\frac{T_h}{s}=1$ in the Movie domain, $\alpha=0.2$, $\beta=1$, $T_b=1$, $\frac{T_b}{s}=1$, $T_h=3$, and $\frac{T_h}{s}=1$ in the Music domain. In $\textit{Book}~$\&$~\textit{Music}$, $\alpha=0.1$, $\beta=1$, $T_b=1$, $\frac{T_b}{s}=1$, $T_h=2.4$, and $\frac{T_h}{s}=1$ in the Book domain, $\alpha=0.1$, $\beta=1$, $T_b=1$, $\frac{T_b}{s}=1$, $T_h=1.8$, and $\frac{T_h}{s}=1$ in the Music domain. In all CDR scenarios, the best ODE solver for the smoothing process is the Euler method, and for the sharpening process is the RK4 method.

\begin{figure}[t]
\setlength{\abovecaptionskip}{0.cm}
	\begin{center}
        \subfigure
        {\begin{minipage}[b]{.32\linewidth}
        \centering
        \includegraphics[scale=0.265]{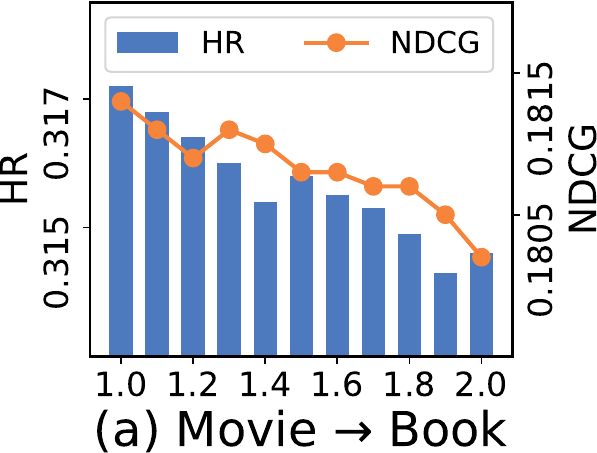}
        \end{minipage}}
        \subfigure
        {\begin{minipage}[b]{.32\linewidth}
        \centering
        \includegraphics[scale=0.265]{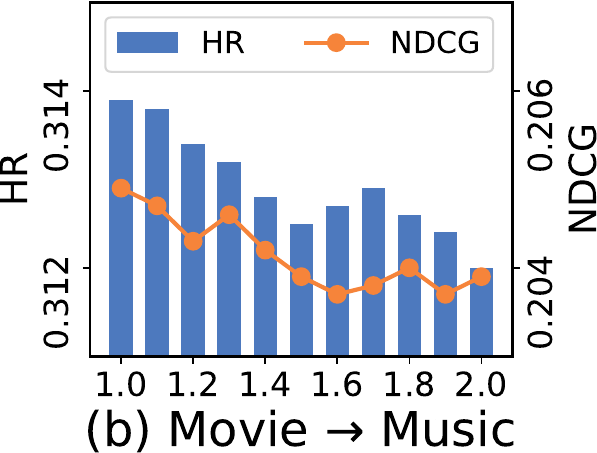}
        \end{minipage}}
        \subfigure
        {\begin{minipage}[b]{.32\linewidth}
        \centering
        \includegraphics[scale=0.265]{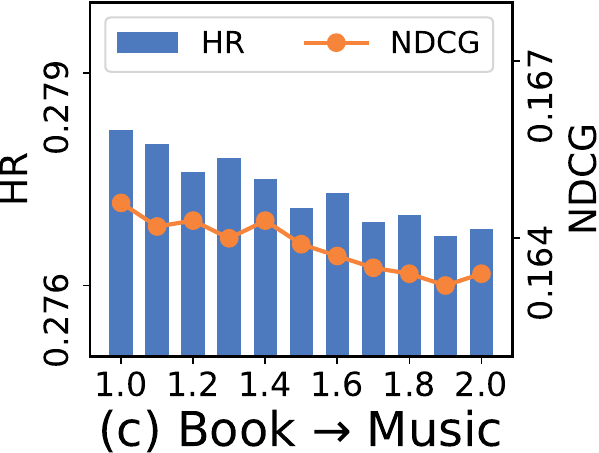}
        \end{minipage}}\\
        \subfigure
        {\begin{minipage}[b]{.32\linewidth}
        \centering
        \includegraphics[scale=0.265]{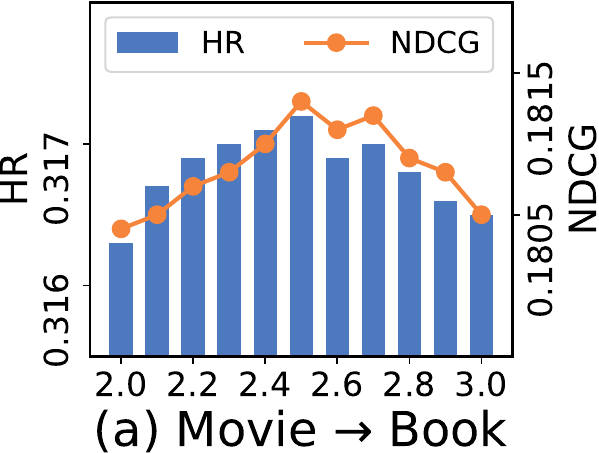}
        \end{minipage}}
        \subfigure
        {\begin{minipage}[b]{.32\linewidth}
        \centering
        \includegraphics[scale=0.265]{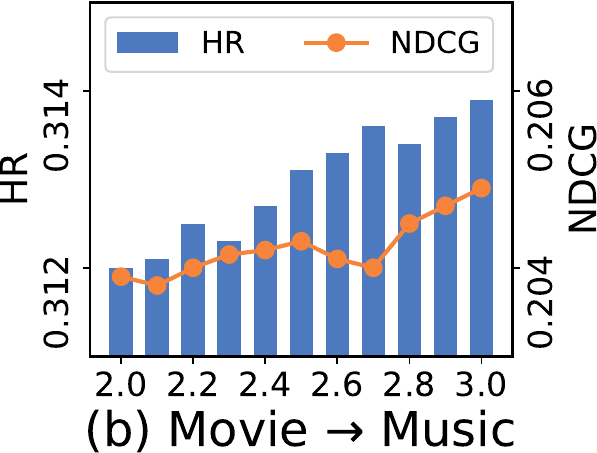}
        \end{minipage}}
        \subfigure
        {\begin{minipage}[b]{.32\linewidth}
        \centering
        \includegraphics[scale=0.265]{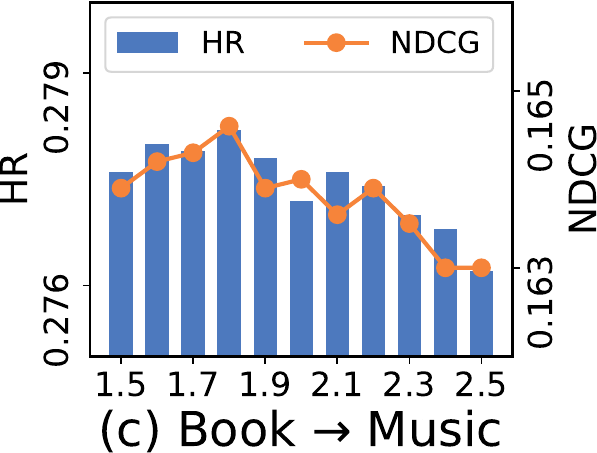}
        \end{minipage}}
        \vspace{0.5em}
        
	\caption{Effect of varying $T_b$ (upper) and $T_h$ (lower).}
	\label{Tb_Th}
	\end{center}
\vspace{-0.9em}
\end{figure}

\begin{figure}[t]
\setlength{\abovecaptionskip}{0.cm}
	\begin{center}
        \subfigure
        {\begin{minipage}[b]{.32\linewidth}
        \centering
        \includegraphics[scale=0.265]{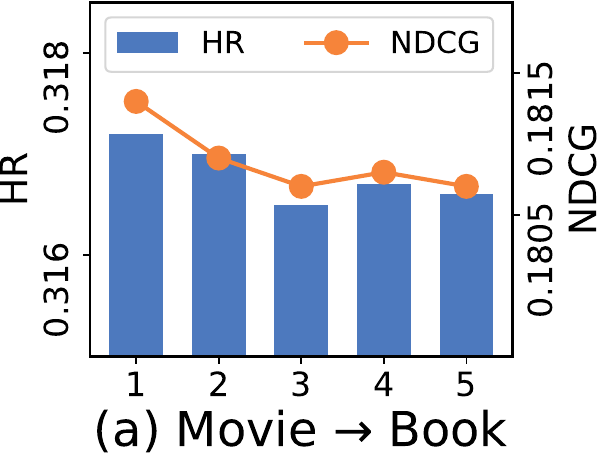}
        \end{minipage}}
        \subfigure
        {\begin{minipage}[b]{.32\linewidth}
        \centering
        \includegraphics[scale=0.265]{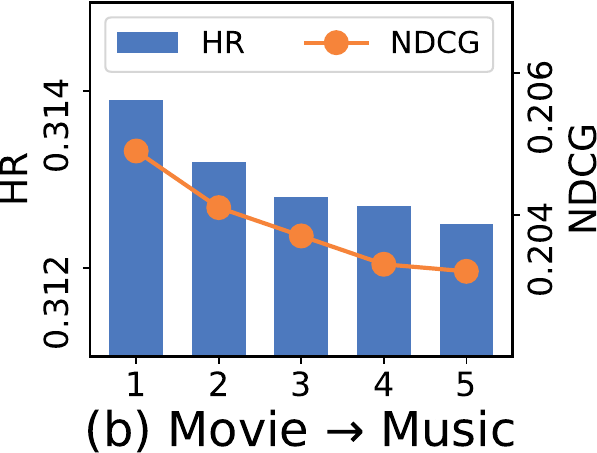}
        \end{minipage}}
        \subfigure
        {\begin{minipage}[b]{.32\linewidth}
        \centering
        \includegraphics[scale=0.265]{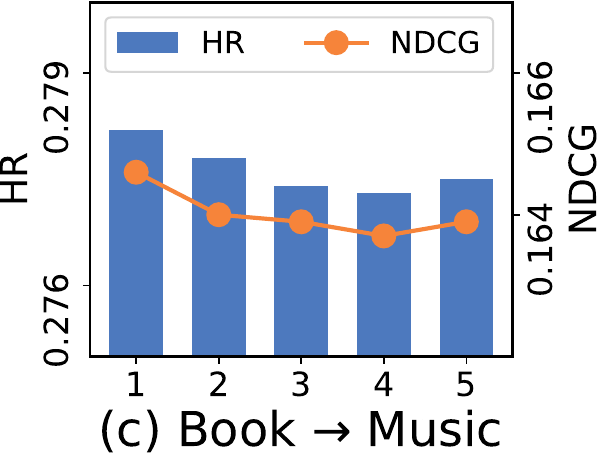}
        \end{minipage}}\\
        \subfigure
        {\begin{minipage}[b]{.32\linewidth}
        \centering
        \includegraphics[scale=0.265]{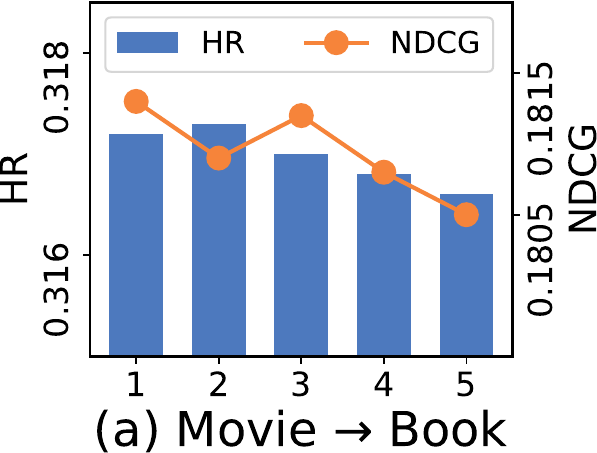}
        \end{minipage}}
        \subfigure
        {\begin{minipage}[b]{.32\linewidth}
        \centering
        \includegraphics[scale=0.265]{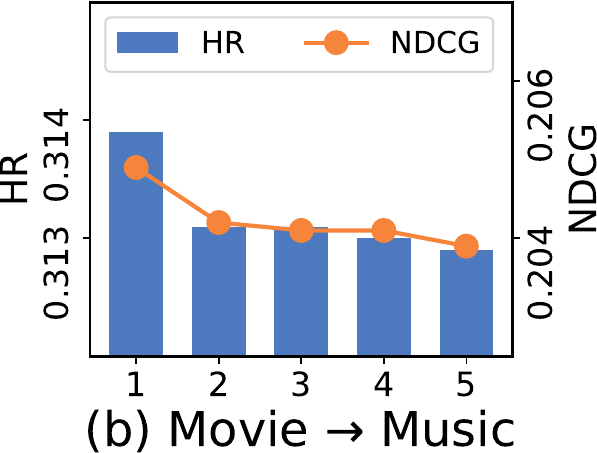}
        \end{minipage}}
        \subfigure
        {\begin{minipage}[b]{.32\linewidth}
        \centering
        \includegraphics[scale=0.265]{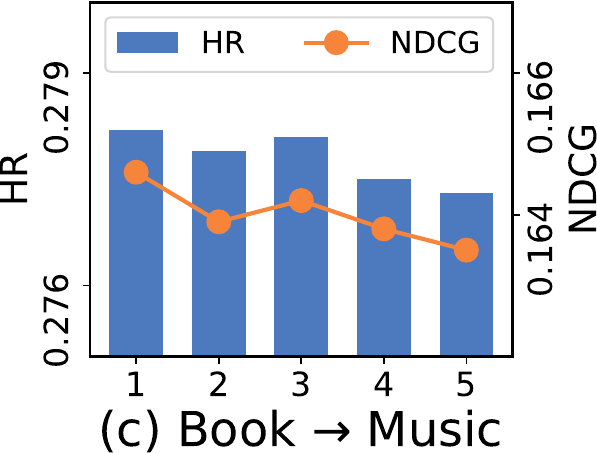}
        \end{minipage}}
        \vspace{0.5em}
        
	\caption{Effect of varying $\frac{T_b}{s}$ (upper) and $\frac{T_h}{s}$ (lower).}
	\label{Tbs_Ths}
	\end{center}
\vspace{-0.9em}
\end{figure}

\subsection{Hyper-parameter Analysis}
We analyze the impact of different hyper-parameter settings in S$^2$CDR, including
the terminal time $T_b$ and $T_h$, the number of steps $\frac{T_b}{s}$ and $\frac{T_h}{s}$. Notably, we report the results for cold-start users in the target domain on three CDR scenarios, while the results in the source domain with similar observations are omitted to save space.

\subsubsection{\textbf{Effect of $T_b$ and $T_h$}}
We investigate the sensitivity of S$^2$CDR to the terminal time $T_b$ of the smoothing process and $T_h$ of the sharpening process. From the results shown in Figure~\ref{Tb_Th}, we observe that all three CDR scenarios achieve the best performance with $T_b=1$, which demonstrate a small value of $T_b$ in the smoothing process can capture the correlations between items and filter out the noise information. Another observation in Figure~\ref{Tb_Th} is that $T_h$ can be set as a large value in all three CDR scenarios to describe user's personalized preference. However, the performance of S$^2$CDR decreases after a certain point in Movie $\to$ Book and Book $\to$ Music. From the perspective of CF, it is attributed to applying the sharpening process too much, since the sharpening process emphasizes user's personalized preference rather than collaborative information. Therefore, we should carefully adjust the value of $T_h$.

\subsubsection{\textbf{Effect of $\frac{T_b}{s}$ and $\frac{T_h}{s}$}}
By varying the number of steps $\frac{T_b}{s}$ and $\frac{T_h}{s}$ for ODE solvers in the smoothing and sharpening processes, we explore how S$^2$CDR performs in Figure~\ref{Tbs_Ths}. We observe that there's no need to use many steps to solve the integral problem of the smoothing and sharpening functions, which makes our method computationally efficient.

\end{document}